\documentclass{aa}  
\usepackage{graphicx}
\usepackage{xspace}
\usepackage{txfonts}
\usepackage{upgreek}
\usepackage[colorlinks=true,linkcolor=blue,citecolor=blue]{hyperref}
\usepackage{booktabs,tabularx,makecell}
\usepackage[table]{xcolor}   % for header color
\usepackage{siunitx}         % for units, optional
\usepackage{hyperref}        % for links
\definecolor{headerblue}{RGB}{8,91,117}
\usepackage{placeins}
\usepackage{stfloats}
\usepackage{natbib}
\usepackage{gensymb}

\begin{document}

   \title{Luminous red novae as shock-powered transients}%: \\ Shock interaction revealed after the cessation of H$^-$}
   \subtitle{I. Electron-scattering wings and deviations from Case~B}
    
    \titlerunning{Electron-scattering wings in LRN}

    \author{Albert Sneppen\inst{\ref{addr:DAWN},\ref{addr:jagtvej}}, Kenta Hotokezaka\inst{\ref{addr:RESCEU}} and Christopher M. Irwin\inst{\ref{addr:Tohoku}}
    }

    \institute{Cosmic Dawn Center (DAWN)\label{addr:DAWN}
    \and
    Niels Bohr Institute, University of Copenhagen, Blegdamsvej 17, K{\o}benhavn 2100, Denmark\label{addr:jagtvej} 
    \and
    Research Center for the Early Universe, Graduate School of Science, The University of Tokyo, Bunkyo, Tokyo 113-0033, Japan\label{addr:RESCEU}
    \and
    Astronomical Institute, Tohoku University, Aoba, Sendai 980-8578, Japan\label{addr:Tohoku}
    }
   
   \date{Received \today; accepted }

% \abstract{}{}{}{}{} 
% 5 {} token are mandatory
 
    % Abstract with 5 {} tokens
    \abstract
    {Luminous red novae (LRNe) are a class of optical transients resulting from the mergers of binary stars or common envelope events. The population displays heterogeneous light curves with extended luminosity plateaus and/or secondary peaks, suggested to be powered by hydrogen recombination or shocks between pre- and post-merger ejecta. However, much of their spectroscopic behaviour remains unexplained. Using a sample of six LRNe, we identify telltale spectral evidence for shock processes across the LRNe luminosity range ($10^{38}$-$10^{41}\,\mathrm{erg/s}$): (i) fast ejecta sweeping up slower upstream material; (ii) composite epochs with two distinct components, namely a cool stellar-like continuum underneath a hot nebular recombination region; and (iii) extremely broad line wings extending to $1000$-$10\,000\,\mathrm{km/s}$. Such line profiles reveal photons scattering off hot electrons in an outflow ($T_e\sim5000$-$10\,000 \,\mathrm{K},\,v_{es}\!\sim\!300$-$500\, \mathrm{km/s}$), with recombination seen from both the upstream and downstream of the shock surface. 
    Additionally, constraints on the density of the upstream medium are obtained from recombination-line ratios, where brighter LRNe show deviations from Case~B, consistent with denser surrounding environments. Given the outflow's velocity and surrounding density constraints, the shock luminosity is sufficient to account for the total energetics of LRNe without requiring other energy-injection mechanisms.
    During the plateau phase, the scattering wings are hidden and line ratios are affected by radiative transport effects, which highlights the utility of post-plateau phase spectroscopy. \newline 
    }
    \maketitle
%
%________________________________________________________________
\section{Introduction}

Interacting binaries are common \cite[e.g.][]{Sana2012} with notable implications for transient observations \cite[e.g.][]{Soker2003,Soker2006,Smith2011}. In particular, a class of optically discovered sources known as `luminous red novae' (LRNe) are believed to originate from stellar mergers. These events are characterised by peak luminosities of $L \!\sim\! 10^{38}$–$10^{41}\, \mathrm{erg\,s^{-1}}$, with redder colours than classical novae and durations ranging from weeks to several months \citep[e.g.][]{Martini1999,Munari2002,Kulkarni2007,Blagorodnova2021,Pastorello2019,Pastorello2023}. The object that established the defining link between LRNe and merging binary stars was the Galactic transient V1309\,Sco \citep{Mason2010}. This system was observed as a contact binary with a decreasing orbital period before undergoing a dramatic brightening of 10 magnitudes \citep{Tylenda2011}. The gradually rising light curve over hundreds of orbital periods preceding final coalescence suggests substantial mass loss \citep[seen in this and other LRNe; e.g.][]{Blagorodnova2017,Metzger2017} similar to the outflow expected from hydrodynamic models \citep[][]{Nandez2014,Pejcha2014,Pejcha2017,MacLeod2018,MacLeod2020,Shroder2020}. 

Final binary coalescence results in an abrupt rise to a luminosity peak, perhaps reminiscent of shock breakout. The initial thermal energy of hot ejecta may power this early luminosity peak \citep{MacLeod2017}, while the subsequent longer-lived plateau is likely sustained by radially distributed heating sources, such as shock interactions \citep{Metzger2017}, hydrogen recombination in various shells \citep{Ivanova2013,Matsumoto2022}, and accretion-powered jets \citep[]{Soker2020}. In the shock-powered interpretation, the energetics result from the collision between rapid post-merger ejecta and pre-existing circumstellar material (CSM), where the CSM is dominated by a toroidal outflow from pre-merger mass loss. Observationally, the spectral energy distribution (SED) during the plateau-phase resembles a thermal blackbody with temperatures in the range $3000 \,\mathrm{K}\lesssim T \lesssim 6000 \,\mathrm{K}$, while photospheric radii undergo a significant expansion ($v_{ph}\!\sim\! 200 - 500 \,\mathrm{km/s}$ from the time-evolution of the photospheric radius) \citep[e.g.][]{Tylenda2011,Pastorello2019,Pastorello2023}. Above this expanding surface, low-velocity ($v < 100\,\mathrm{km/s}$) absorption features are often observed, arising in the surrounding CSM from pre-merger ejecta \citep[e.g.][]{Mason2010, Mason2022}. This material can be substantial for the brightest LRNe, as evidenced by the presence of weak absorption lines, including forbidden transitions of Ca\,\textsc{ii} (\(3p^6 4s - 3p^6 3d\)) as seen in AT2018bwo \citep{Blagorodnova2021}. Finally, from a few weeks to several hundred days after the merger, the LRN optical light curve decays abruptly; the spectral continuum changes dramatically with a peak in near-infrared (NIR) \citep[e.g.][]{Tylenda2016,Rudy2025} and is scarred by molecular absorption features \citep[e.g.][]{Martini1999,Kaminski2015}.

\begin{figure*}
    \centering
    \includegraphics[width=\linewidth, viewport=42 22 860 530 ,clip=]{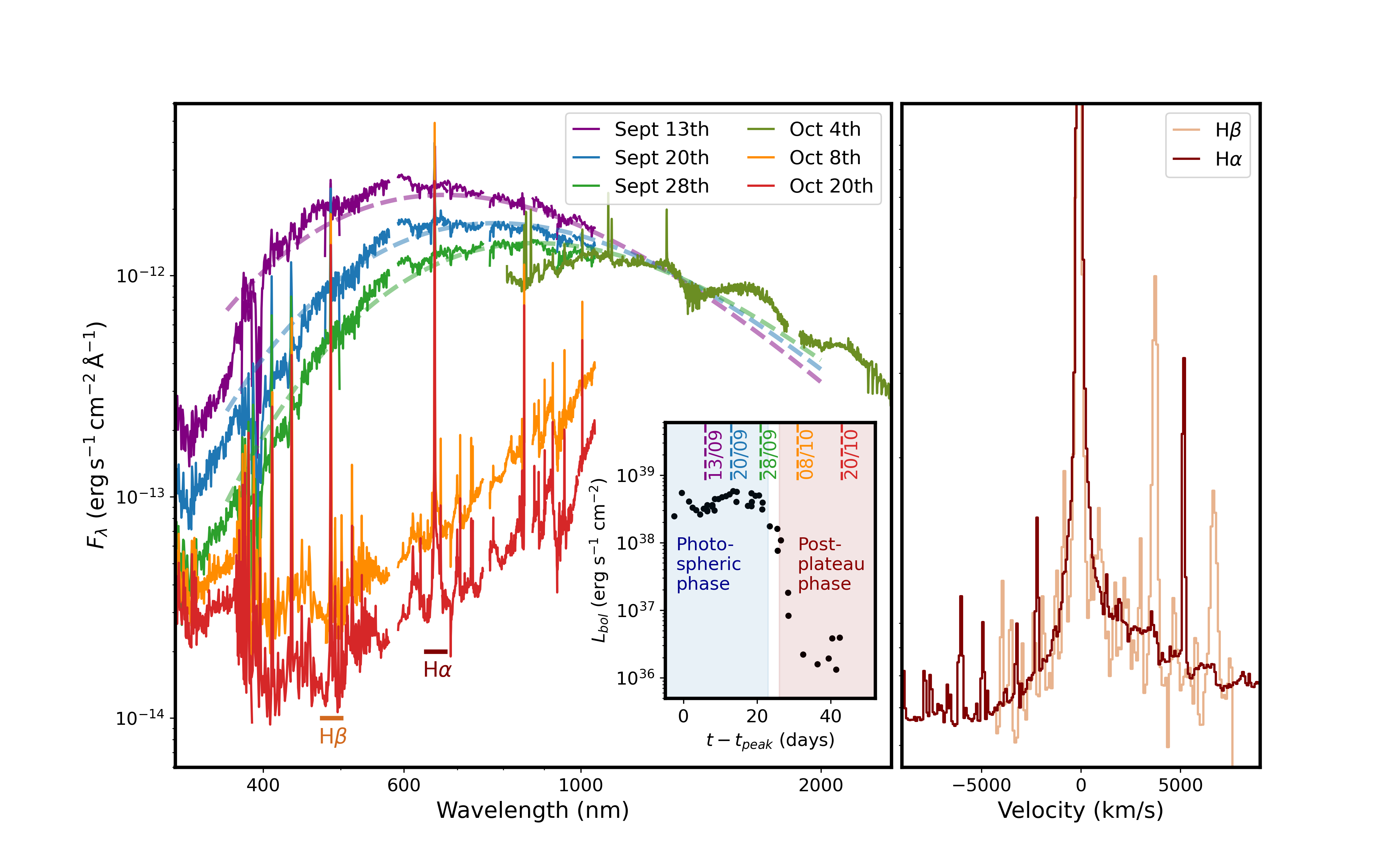}
    \caption{Spectra of the LRN V1309\,Sco illustrating the spectral change from photospheric to post-plateau epochs \citep{Mason2010,Rudy2025}. Inset panel: Bolometric luminosity evolution derived from blackbody fits to photometry (see Appendix~\ref{sec:BB_fit}). Right panel: H$\alpha$ and H$\beta$ lines during the post-plateau epoch (scaled to the Case~B recombination-line ratio and corrected for dust extinction), showing narrow ($v\sim50\,\mathrm{km/s}$) and very broad components ($v\gg1000 \mathrm{km/s}$).
    Blackbody functions reproduce the early SED until $T_{BB} \!\sim\! 3000\,$K, at which point an abrupt luminosity drop occurs. At this stage, the continuum becomes reddened and non-blackbody-like, with prominent free-bound emission, recombination lines, and broadened H$\alpha$ and H$\beta$ features, identified here as electron-scattering wings.  }
    \label{fig:Fig1}
\end{figure*}

\begin{figure*}
    \centering
    \includegraphics[width=\linewidth, viewport=42 22 860 530 ,clip=]{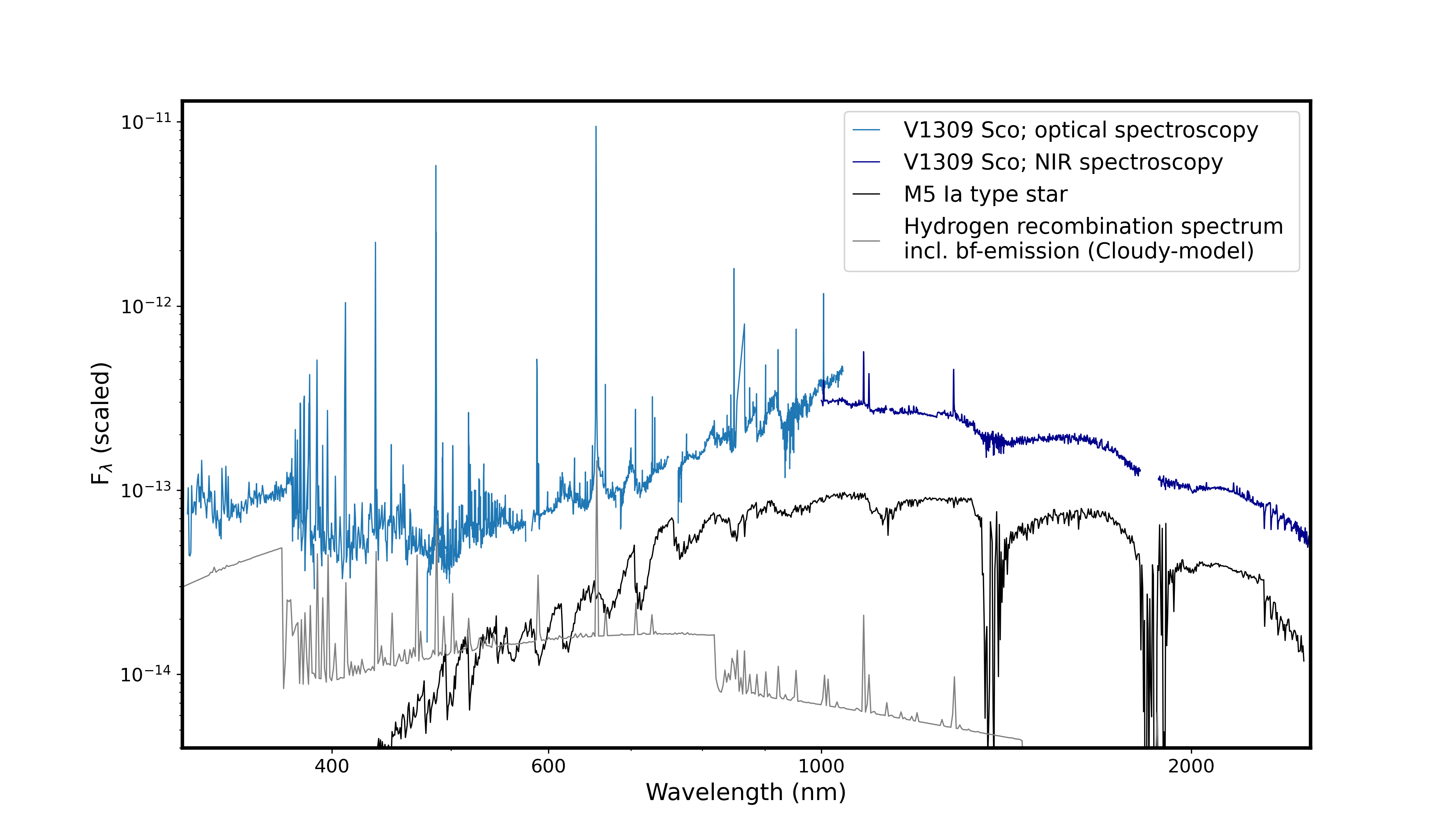}
    \caption{ V1309\,Sco optical and NIR spectroscopy in post-plateau epochs compared to i) a model of recombination emission producing free-bound emission and recombination lines and ii) an M-star type spectrum producing the red continuum. The NIR spectrum was taken four days earlier than the optical (i.e. at a brighter phase) and is therefore scaled down by a factor of 5. 
    Matching the red optical or the NIR continuum with the M-type star template would suggest a corresponding surface radius of $3(\pm1)\cdot10^3R_{\odot}$, which aligns with the expectation from blackbody radii in photospheric epochs (see Appendix~\ref{fig:lbol}). Associating the recombination lines and the observed spectral continuum at $\lambda\!<\!500\,\mathrm{nm}$ with a hot ionised environment suggests that this component contributes $\!\sim\!$20\% of the bolometric luminosity. }
    \label{fig:mstar}
\end{figure*}

As we argue in this paper, the sudden transition into the post-plateau regime permits a direct view through the previously reprocessing photosphere into the LRNe energy source.
In Sect.~\ref{sec:sample}, we summarise the sample of analysed LRNe. In Sect.~\ref{sec:two_comp}, we identify two distinct SED components in post-plateau phase spectra with a coupled temporal evolution. 
In Sect.~\ref{sec:electron-scattering}, we highlight the physical constraints provided by the electron-scattering wings found in post-plateau epochs, which show telltale signatures of shock interaction from a thermal broadening of hydrogen lines to widths of order $v=10^3-10^4\,\mathrm{km/s}$.  
In Sect.~\ref{sec:Density}, we discuss using recombination-line ratios to derive example constraints on the upstream density, which in combination with the velocity of electron-scattering wings suggest a shock luminosity is sufficient to power LRNe (even for low-density and low-luminosity objects). In a companion paper (Paper~II), we show how this shock interpretation motivates a description of the light curve evolution as a CSM shock breakout followed by a luminosity plateau from continued interaction, similar to the case of type IIn supernovae. This framework entails notable predictions for the CSM density structure and, by extension, the temporal evolution of the broad electron-scattering wings identified here.

\begin{figure*}
    \includegraphics[width=\linewidth]{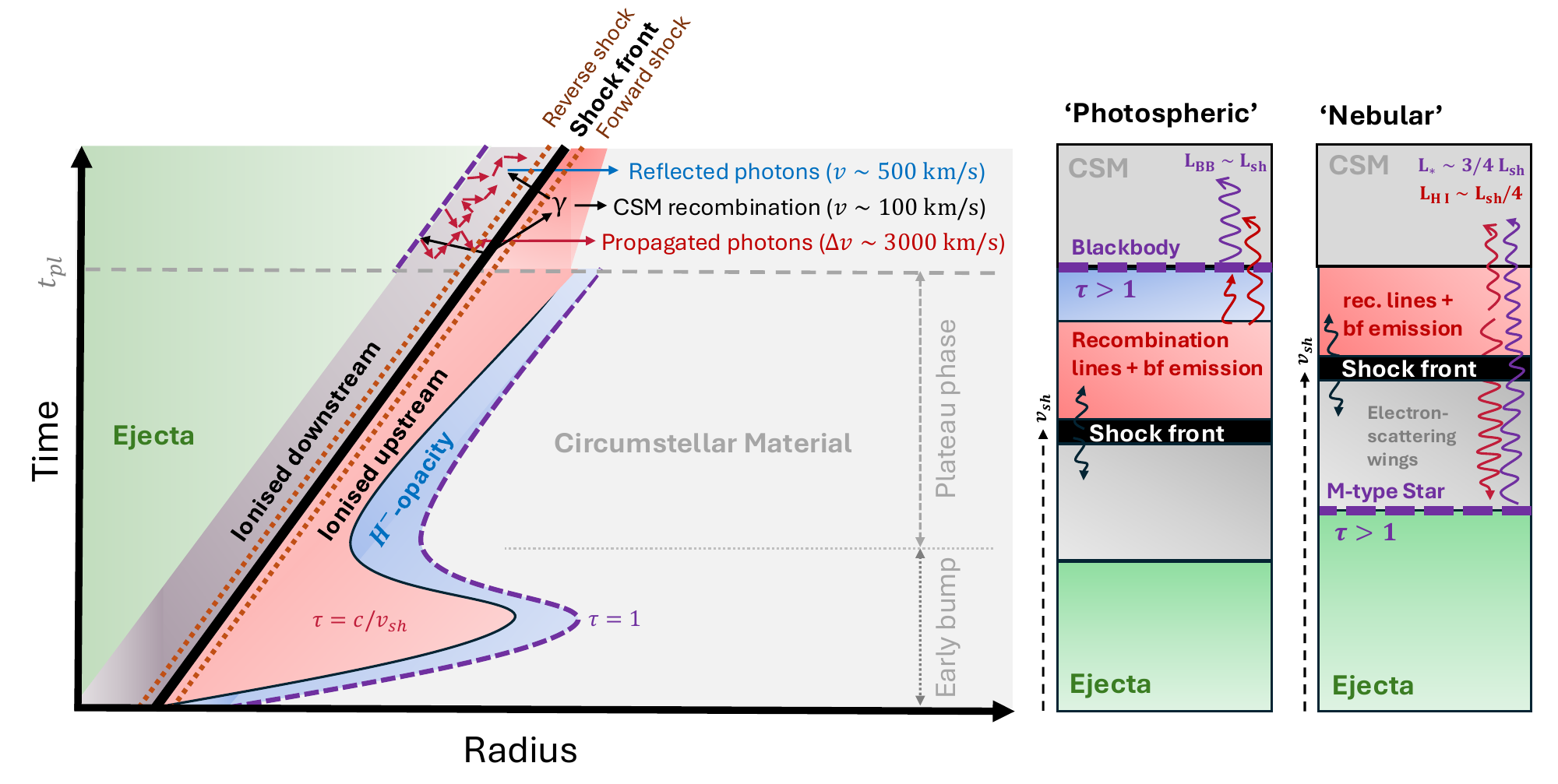}
    \caption{Left: Schematic illustration of the various shells as a function of time and radius for the proposed interpretation. Right: Radial structure in photospheric and post-plateau epochs, where the key difference is whether the optically thick $\tau=1$ surface is ahead of or behind the shock front. 
    As the shock front moves outwards, the CSM immediately upstream is ionised (producing strong recombination lines and free-bound emission), while in photospheric epochs an H$^-$ opacity layer reprocesses the shock emission to a thermal blackbody. Once the upstream material becomes optically thin, the recombination layers are directly exposed, while transport through and scattering within the hot ionised downstream region produce the observed asymmetric electron-scattering wings. Multiple scattering generates broad wings with $\Delta v\!\sim\!3000\,{\rm km/s}$, even though no ejecta component reaches such velocities. 
    The M-type stellar NIR spectrum arises from the ejecta surface, which is irradiated by photons from the shock. Because both the recombination lines and the NIR stellar component are powered by the same irradiation source, their relative luminosities remain approximately constant.
    }
    \label{fig:shematic_electron_scattering}
\end{figure*}

\begin{figure*}
    \includegraphics[width=\linewidth]{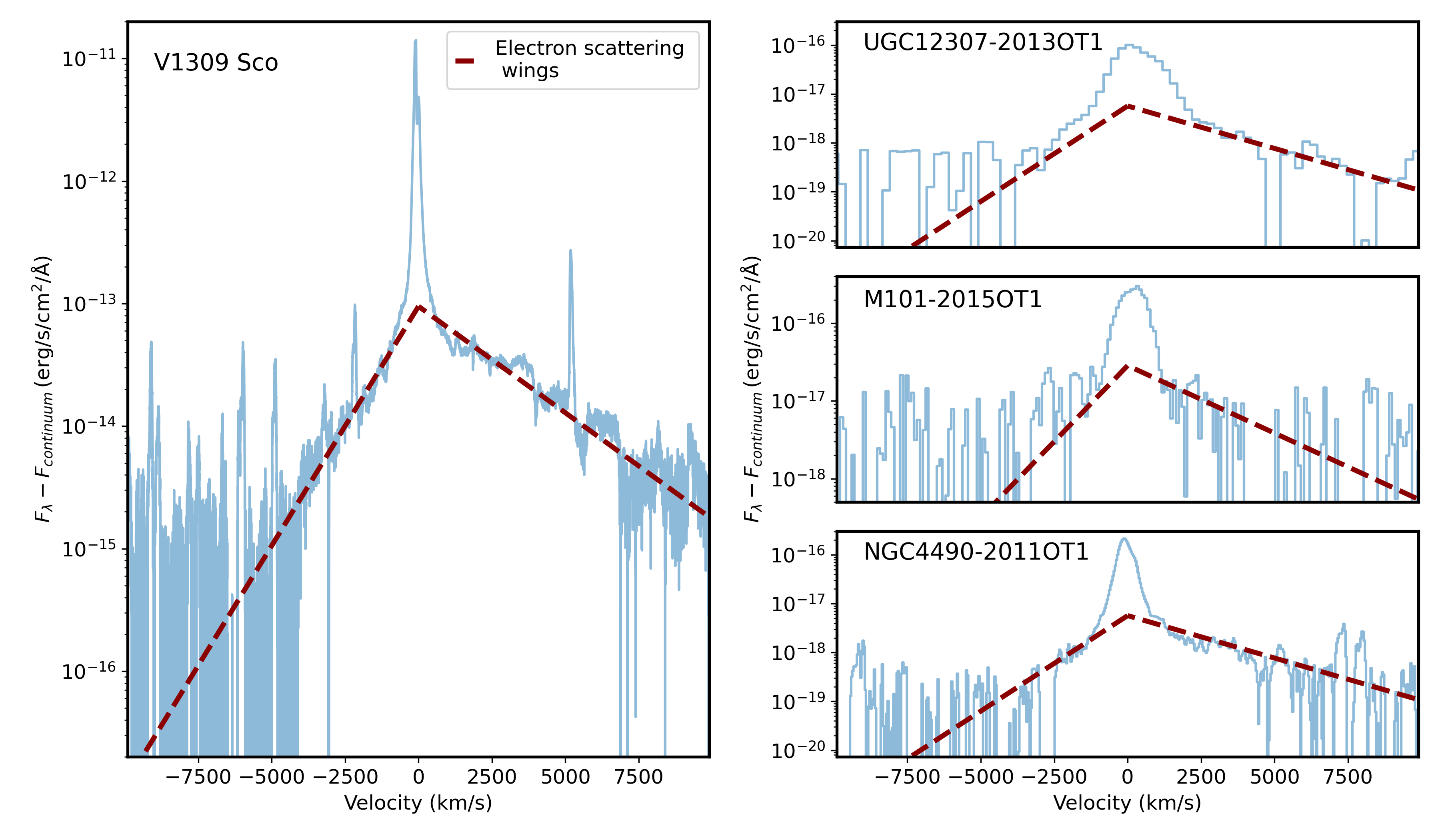}
    \caption{H$\alpha$ line profiles for four LRNe -- V1309\,Sco \citep{Mason2010}, M101-2015OT1 \citep{Goranskij2016}, NGC4490-2011OT1, and UGC12307-2013OT1 \citep{Pastorello2019} -- taken in early post-plateau epochs (see light curves in Fig.~\ref{fig:lbol}). They all show extended asymmetric wings consistent with exponentially declining intensity away from the line centre. The dashed red lines show the electron-scattering wings for an optically thick outflow with $v_{es} \!\sim\! 500$\,km/s and $v_{th} \!\sim\! 500$\,km/s, which for these objects produces a decent match to the broad components. }
    \label{fig:Halpha_wind}
\end{figure*}

\section{Luminous red novae sample}\label{sec:sample}
Our sample comprises six well-studied LRNe -- V1309\,Sco, V838 Mon, M101-2015OT1, NGC 4490-2011OT1, UGC 12307-2013OT1, and AT2021blu -- chosen to span the observed diversity of light-curve morphologies while providing the spectroscopic coverage necessary for detailed constraints. Their light-curve evolution and the key spectral signatures during the plateau and post-plateau phases are summarised in Appendix~\ref{sec:BB_fit}, while Fig.~\ref{fig:Fig1} presents a spectral sequence of one of these landmark LRNe, V1309\,Sco, across plateau and post-plateau epochs. Throughout this work, we use the term `photospheric phase' to describe the bright, extended phase during which the continuum is similar to a blackbody, indicating efficient reprocessing. By contrast, as seen in Fig.~\ref{fig:Fig1}, we use the term `post-plateau phase' for later spectra in which a cool stellar-like optical/NIR component, often accompanied by molecular bands, coexists with a hotter ionised-gas component producing hydrogen recombination continuum and emission lines. We use `nebular' only to refer to this hotter recombination component.

V1309\,Sco is used here as a key reference object: among all observed LRNe, it has the highest signal-to-noise ratio (S/N) spectroscopic coverage across the initial peak, luminosity plateau, and particularly in the immediate post-plateau phase. Accordingly, in Sects.~\ref{sec:two_comp} and~\ref{sec:electron-scattering}, we adopted it as our fiducial event for post-plateau studies of the two components of emission and the electron-scattering wings, respectively. This object is, however, not representative of most of the LRN population, which deviates from Case~B recombination. As discussed in Sect.~\ref{sec:Density}, such deviations are more typical of LRNe such as V838 Mon, M101-2015OT1, and AT2021blu, which we show are embedded in denser circumstellar environments. This may be the physical cause of their brighter nature.

\section{Two energetic components in the post-plateau spectra: Blue nebular emission and a NIR star }\label{sec:two_comp}

While both bound-free features and hydrogen recombination lines are superimposed on the blackbody in photospheric epochs, the transition to the post-plateau phase is marked by a fractional and an absolute increase in the luminosity of H$\alpha$, suggesting a less reprocessed version of the original emission. In these post-plateau epochs, which probe deeper into the ejecta, several interesting spectral phenomena emerge, as illustrated in Fig.~\ref{fig:mstar}. 
First, prominent bumps of free-bound emission prominent bumps of free-bound emission, such as a Balmer jump, and a plethora of recombination lines, including H$\alpha$-H$_{22}$, and P$\alpha$-P$_{29}$ for V1309\,Sco indicate a hot ionised environment. This is qualitatively similar to the Balmer and recombination emission seen at blue wavelengths in T Tauri stars, where it is interpreted as evidence of shock-powered processes \citep{Basri1989,Calvet1998}. 
Second, at near-infrared wavelengths, a stellar-like spectrum dominates, which, across LRNe, evolves in time towards cooler stellar spectral types \citep[e.g.][]{Martini1999,Tylenda2005} and displays stellar radii comparable to the blackbody radii of previous photospheric epochs. 

For V1309\,Sco, the observed hot ionised emission (free-bound and recombination lines) is $\!\sim\!1/5$ of the total bolometric luminosity, making it a significant, although sub-dominant, contribution. Figure~\ref{fig:mstar} illustrates that this  $\!\sim\!1/5$ luminosity can be extracted either from the relative intensity of the \textsc{cloudy} and NIR-star model or from dividing the observed spectrum into its nebular-dominated ($\lesssim500\,$nm) and stellar-dominated component ($\gtrsim500\,$nm).
While both the hot ionised emission and the red stellar luminosity decrease by 40\% between 8 and 20 of October, the fractional luminosity of the hot ionised component remains constant. We also traced individual recombination lines over a longer post-plateau temporal baseline. For instance Pa$\delta$ and Pa$\gamma$ fade by 70\% in V1309\,Sco but are constant relative to the bolometric luminosity. This means that Pa$\delta$ and Pa$\gamma$ are brighter in the immediate post-plateau spectrum compared to the photospheric phase, i.e. these photons escape without being reprocessed into the cool continuum. Similarly, H$\alpha$ decreases by over 80\% in NGC4490-2011OT1 but remains constant relative to the continuum (see~Fig. \ref{fig:lbol}). A similar coherent fading is also seen for the continuum and the weak high ionisation lines (see Sect.~\ref{sec:he4686}).
Ultimately, prominent recombination lines are a robust feature of LRNe from photospheric through early post-plateau epochs (e.g. Figs.~\ref{fig:AT2021blu_spectra} and~\ref{fig:V838Mon}), and this coherent fading of the recombination lines and NIR stellar continuum is thus seen in both the LRNe with multiple post-plateau epoch spectra (see Figs.~\ref{fig:Fig1} and \ref{fig:NGC4490}). This spectroscopic decomposition highlights two distinct radiative-transfer components that are nevertheless intrinsically linked energetically, fading at the same rate. 

We propose that these two post-plateau spectral energy components fade coherently, because they are illuminated by the same shock front (see Fig.~\ref{fig:shematic_electron_scattering}). In this interpretation, the transition from photospheric to post-plateau could represent the $\tau=1$ surface receding over the shock front. 
Before the shock front becomes directly visible, due to optical depth effects i) the relative strength of lines to continuum can change across epochs, and ii) the blackbody radius can increase in the initial luminosity peak before relaxing again.
In post-plateau epochs, upstream of the shock, the ionised and slowly expanding CSM produces narrow recombination lines and bound-free emission, while the dense downstream produces an optically thick stellar surface illuminated from above. The latter environment is also characterised by a denser, ionised, and more rapidly expanding outflow, providing suitable conditions for producing asymmetric electron-scattering wings, to which we now turn.

\section{Electron-scattering wings}\label{sec:electron-scattering}

Particularly puzzling LRNe features include the observed width of H$\alpha$ as previously noted in V1309\,Sco by \cite{Mason2010} and also seen in the post-photospheric phase of more luminous LRNe \citep[M101-2015OT1, NGC4490-2011OT1, UGC12307-2013OT1, AT2018hso in][]{Goranskij2016,Blagorodnova2017,Pastorello2019,Cai2019}. These wings are shown for a few objects in Fig.~\ref{fig:Halpha_wind} and naively display extremely large velocities, one to two orders of magnitude beyond those inferred from the photospheric velocities or the emission or absorption line-widths in any epoch. In this analysis, we suggest that these are the spectral signatures of electron-scattering wings from a hot population of outflowing electrons, shock-heated by the collision of post-merger ejecta with the CSM formed by pre-merger ejecta. This is analogous to the systems analysed in \cite{Laor2006,Huang2018}, and it has been noted as a possible, though untested, explanation for a very tentative broad component in M101-2015OT1 \citep{Goranskij2016}. We argue that several physical considerations support this interpretation: 

\begin{enumerate}
    \item The apparent velocity is far beyond the typical escape velocity of LRN binary-progenitors, whereas repeated thermal broadening from an optically thick environment provides a natural pathway to large velocities.
    \item The feature is not obvious in the early photospheric phase, which may argue against its originating outside the slower photosphere. Thermal broadening, in contrast to Doppler broadening, provides a natural explanation for why the broader component originates in the inner ejecta. However, we note that the order-of-magnitude decrease in continuum flux from the photospheric to the post-plateau phase, and the factor-of-few increase in the H$\alpha$ line luminosity make it difficult to identify the broad component against the continuum.
    \item The broad component is observed in both H$\alpha$ and H$\beta$, and the prominence of the broad component relative to the narrow is similar for both.
    Broadening by electron scattering naturally predicts such a wavelength-independent effect on all sufficiently prominent lines. 
    \item The broad components show blue--red asymmetry. Raman scattering could produce sufficiently broad lines for rather large column-densities $\log(N_H)>21$ and asymmetric lines for $\log(N_H)\gtrsim23$. However, using the modelling of \cite{Kokubo2024}, we find that Raman scattering cannot simultaneously reproduce the observed width and asymmetry. While Lorentzian profiles are sometimes used to parametrise these broad LRN features \citep[e.g.][]{Pastorello2019}, it is worth emphasising that such profiles would not produce asymmetric features and would require more pressure broadening than observed even in white dwarfs. In contrast, electron-scattering wings in systems with outflows (e.g. $v\!\sim\!100$-$1000\,\mathrm{km/s}$) must display predictable asymmetry between blue and red wings \citep[e.g.][]{Huang2018}. In particular, the degree of asymmetry implied by the outflow velocity can be tested against the system’s other velocity diagnostics.
    \item The broad wings follow the expected functional form of electron-scattering wings. This exhibits -- as seen in Fig.~\ref{fig:Halpha_wind}, particularly for the high S/N objects we observe -- an approximately exponential decline in intensity with increasing distance from the line centroid over several orders of magnitude in flux and several e-foldings in velocity.
\end{enumerate}

\begin{figure*}
    \includegraphics[width=\linewidth]{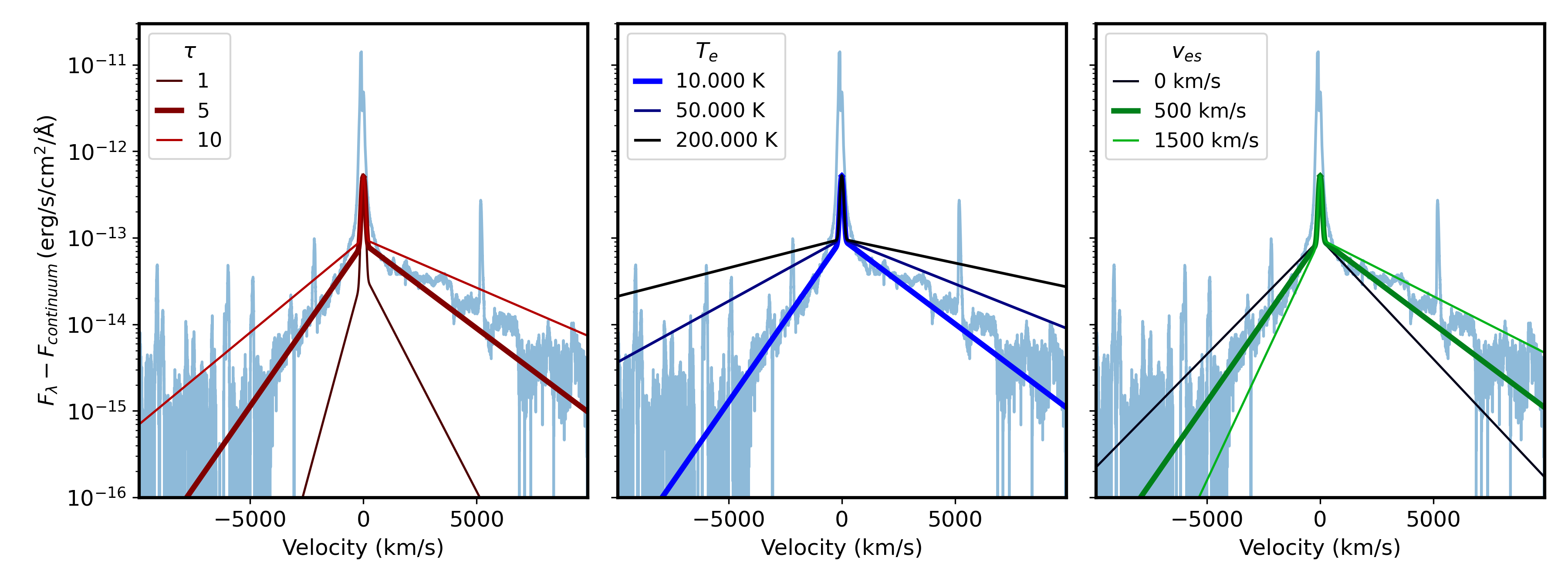}
    \caption{V1309\,Sco H$\alpha$ line and electron-scattering wings around an intrinsically narrow Gaussian line with varying electron-scattering optical depth ($\tau_{es}$), electron temperature ($T_e$), and bulk outflow velocity ($v_{es}$). Temperature and optical depth together set the e-folding width, while the outflow velocity sets the red-to-blue asymmetry. Optical depth also affects the relative prominence of narrow and broad components, but we note the narrow component is likely to be affected by geometric freedom, where recombination photons can escape without passing through the hot ejecta region. For each panel, we fixed the remaining parameters at $\tau_{es}=5, \,T_e=10\,000 \,\mathrm{K}, \,v_{es}=500\,\mathrm{km/s}$ (which we find to be plausible physical parameters; see main text). }
    \label{fig:Electron_scattering_three_panel}
\end{figure*}

\begin{figure*}
    \includegraphics[width=\linewidth]{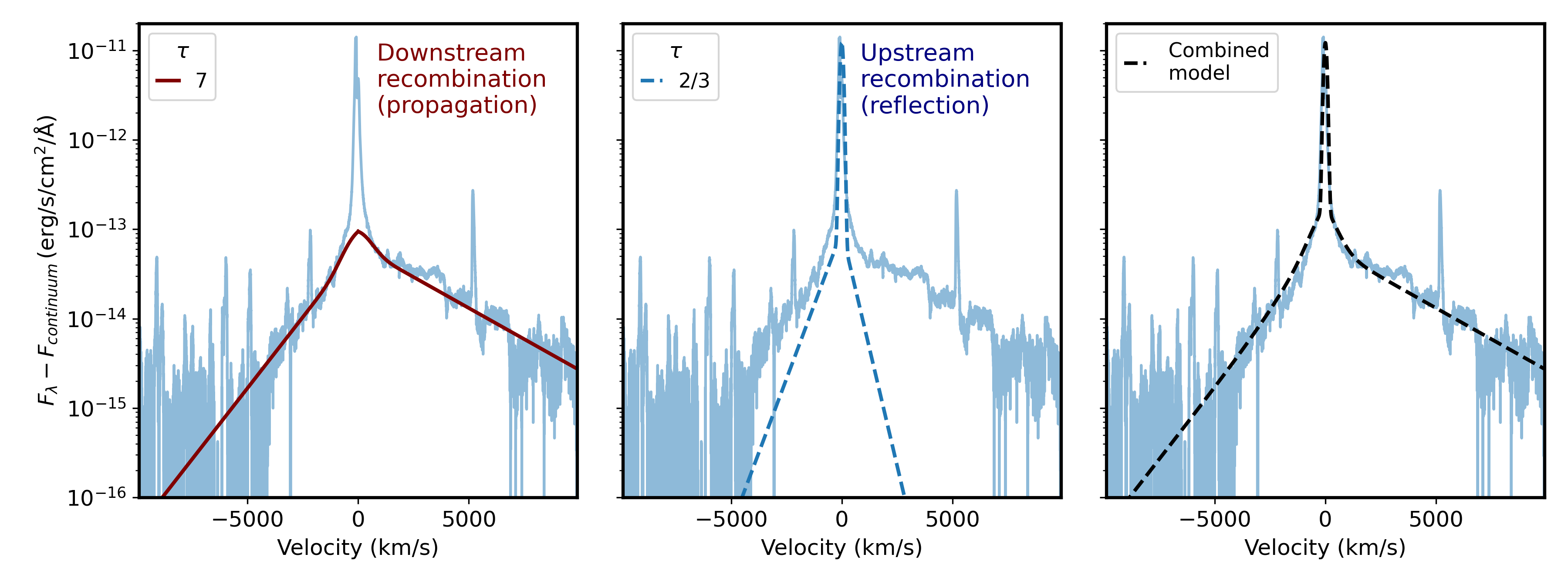}
    \caption{V1309\,Sco H$\alpha$ line and the two distinct scattering components proposed in this interpretation. Left panel: Recombination-line photons produced in the shocked ejecta. These travel through the optically thick electron-scattering region, producing a broad ($\!\sim\! 1000-10\,000$ km/s) exponential, redward-skewed feature with a strongly diminished unscattered core. Central panel: Most recombination photons upstream of the hot electron region escape undisturbed. A minority propagates inwards, reflects off the hot electrons, then escapes without further interaction, producing a bright narrow line together with a typically single-scattered, blueward biased component. This single-scattered contribution does not require additional parameters, since it arises from the same temperature and bulk velocity as the forward-scattered component, while further scatterings are exponentially suppressed. Right panel: Combined effects of upstream and downstream electron scattering produce a very narrow line ($\lesssim100\,\mathrm{km/s}$), a broader core ($\!\sim\!300\,\mathrm{km/s}$), and a  redward-skewed, multiple-scattered and extremely broad component ($\!\sim\!1000-10\,000\,\mathrm{km/s}$). }
    \label{fig:Electron_scattering_three_panel_upstream_downstream}
\end{figure*}

\subsection{Constraints from electron-scattering wings}\label{sec:electron_constraints}
We modelled the electron-scattering wings using the framework established in \cite{Auer1972,Laor2006,Huang2018} \footnote{We note, the analytical frameworks in these studies assume spherical symmetry, which we generalise in Appendix~\ref{app:equatorial_scattering} to a cylindrical symmetry. This introduces further degeneracy in inferring the optical depth from the escape fraction but ultimately does not change the fundamental observables.}.
Photons escaping from the ejecta without scattering on hot electrons should be observed as an intrinsically narrow component. Conversely, the broad exponential component with $\Delta v\!\sim\!1000-10\,000\,\mathrm{km/s}$ represents the multiply scattered photons. 
If the electrons are in an outflow, the resulting redistribution becomes biased towards redder wavelengths. This is because in the comoving frame of any fluid element the rest of the outflow is either stationary or recedes. Thus, the electron-scattering wings fundamentally permit three important quantities to be measured: i) the electron column density, $N_e$; ii) the hot electron temperature, $T_e$; and iii) the approximate outflow velocity, $v_{es}$. Inferences on the exact temperature may be difficult for complicated geometries given the significant degeneracy between electron column density and temperature in producing the broad components, but the presence of a substantial scattered fraction still places useful constraints on the combinations of $T_e$ and $\tau_{es}$. Lastly, while we do not explore it within this work, the ratio of continuum absorption to scattering opacity sets the loss of photons from the line itself. As the highly scattered photons indicative of the high-velocity tails of the exponentials experience the longest pathways, these are also most likely lost to continuum processes. Thus, the exponential slope can be steepened by sources of continuum absorption opacity (e.g. dust), but the exponential shape is still maintained \citep{Huang2018}. 

We first constrained the electron column density from the optical depth to electron scattering, $\tau_{es}$, by comparing the photons scattered off hot electrons with the unscattered photons (see Fig.~\ref{fig:Electron_scattering_three_panel}, left panel). The conservative lower limit is if the hot electrons entirely surround the region characterised by recombination, such as recombination photons propagating through an ionised CSM upstream of the shock. Then, regardless of the exact radial distribution of the hot electrons, the requirement that only a minority of the line luminosity is in the broad components implies $\tau_{es}\!<\!1$ \cite[e.g. Fig. 7 in][]{Huang2018} (as most photons travelled unabated). Specifically, the optical depth can be approximated given the escape probability, $p_{esc}(\tau_{es})=(1-e^{-\tau_{es}})/\tau_{es}$, which describes the line-intensity ratio of the narrow to the total line: $p_{esc}\!\sim\!0.8-0.9$ for V1309\,Sco, UGC12307-2013OT1, and NGC4490-2011OT1. This would suggest $\tau_{es}\gtrsim0.2\text{-}0.4$. Given $\tau_{es} = \sigma_{T} N_e$, where $\sigma_T$ is the Thomson cross-section and $N_e$ is the number column density of these hot electrons. Therefore, we can deduce that $N_e\gtrsim(3\text{-}6)\times10^{23} \,\mathrm{cm^{-2}}$. In Sect.~\ref{sec:two_comp_es}, we show that the majority of the narrow recombination is likely to take place outside the shocked hot electron region, which makes it difficult to compare the narrow and broad component fluxes (as the broad may only result from a fraction of the narrow flux) and permits $\tau_{es} \gg1$. 

We next constrained the electron temperature from the width of the broad component (see Fig.~\ref{fig:Electron_scattering_three_panel}, central panel). Using the lower limit on $\tau_{es}$ discussed above and using the fitting formulae in \cite{Laor2006}, we can for example constrain $T_e\lesssim200\,000\,$K. However, geometric freedoms can be invoked with some lines of sight to the recombination regions potentially not passing through the hot component implying more broadening from $\tau_{es}$ and a correspondingly lower $T_e$. Particularly, this is motivated by the observational fact that the narrow-to-broad line ratios suggest $\tau_{es}\!\sim\!0.2-0.4$ for all the various objects in Fig.~\ref{fig:Halpha_wind} and for both the post-plateau epochs observed in the case of V1309\,Sco. There is no clear reason for such a similar and somewhat finely tuned $\tau_{es}$ to be present in the diversity of conditions. Furthermore, the highly effective cooling at $T_e\!\sim\!200\,000\,$K \citep[e.g.][]{Sutherland1993,Schure2009} might create an expectation for a diversity of widths for various stages of cooling. In contrast, once cooling becomes inefficient at around $10\,000$\,K, we may adopt this as a lower prior on $T_e$, which given the observed e-folding scale in velocity correspondingly puts an upper limit that $\tau_{es}\lesssim15$ ($N_e\lesssim40 \,\mathrm{g/cm^2}$). For $T_e=10\,000\,\mathrm{K}$, the thermal velocity width, $v_{th}=(2k_BT_e/m_e)^{1/2}$, redistributing photons per scattering is thus of order $550 \,\mathrm{km/s}$, which is comparable to the typical outflow velocity in these systems. If molecules such as $H_2$ provide sufficient cooling, the boundary of efficient cooling could perhaps be lowered somewhat further, so we note $T_e\!\sim\!5000$\,K would lead to characteristic reshuffling velocity of $\!\sim\!380 \,\mathrm{km/s}$. 

We finally constrained the outflow velocity using the dramatic asymmetry between the blue and red wings (see Fig.~\ref{fig:Electron_scattering_three_panel}, right panel). Such asymmetric wings are well-established signatures of outflows \citep{Auer1972} in symbiotic stars \citep{Seker2012}, Wolf-Rayet stars \citep{Hillier1991}, or type II-P supernovae with significant wind velocity \citep{Roming2012}.
Modelling the electron-scattering with a Monte Carlo code analogous to \cite{Huang2018}, we find that the factor of two difference in the velocity e-folding scale in the blue wing compared to the red requires $v_{es}$ to be comparable to $v_{th}$. In the conservative optically thin limit with $T_e\!\sim\!200\,000\,\mathrm{K}$, this requires $\!\sim\!$2000 km/s ejecta, which would still be the largest velocity found in these systems.
However, the velocity becomes more reasonable in the optically thick and lower temperature limit. Here, the blue exponential wing is around twice as steep as the red side (to within 20\%) across a range in velocity $v_{es} = 400-800\,\,(T_e/10\,000\,\mathrm{K})^{1/2} \,\,\mathrm{km/s}$. This may explain why the velocity asymmetry is largely the same in all three LRNe shown in Fig.~\ref{fig:Halpha_wind}: thermal electron velocity, once cooling becomes inefficient, is close to the escape velocity of the stellar progenitors, which is the characteristic ejecta velocity for stellar mergers.
While the velocity from the asymmetry of the wings remains similar to the velocities of the expanding photosphere in photospheric epochs, it is larger than the characteristic off-set velocity of the narrow recombination lines, which move at $v_{rec}=80-100$\,km/s relative to the rest frame \citep[e.g. see Fig. 7 in][]{Mason2010}. Thus, in V1309\,Sco the dominant observed recombination lines occur in a region moving outwards at a modest 100 km/s.

Lastly, we emphasise that the optically thin case requires fine-tuning in several ways, whereas the optically thick electron-scattering wing case naturally explains several observed properties. First, the observed wings in the variety of objects display similar width, because they are optically thick and at a similar boundary of efficient cooling. Second, it alleviates the large velocities needed to explain the asymmetry and naturally predicts that the velocity asymmetry should be similar across objects. Third, the escape-fraction is large and similar across objects and time, because most of the recombination is happening in slow material upstream of the shock. In the following we delve into this last point, which could be an instance of an astrophysical system with multiple electron-scattering components.

\begin{figure}
    \includegraphics[width=\linewidth,viewport=15 15 342 550 ,clip=]{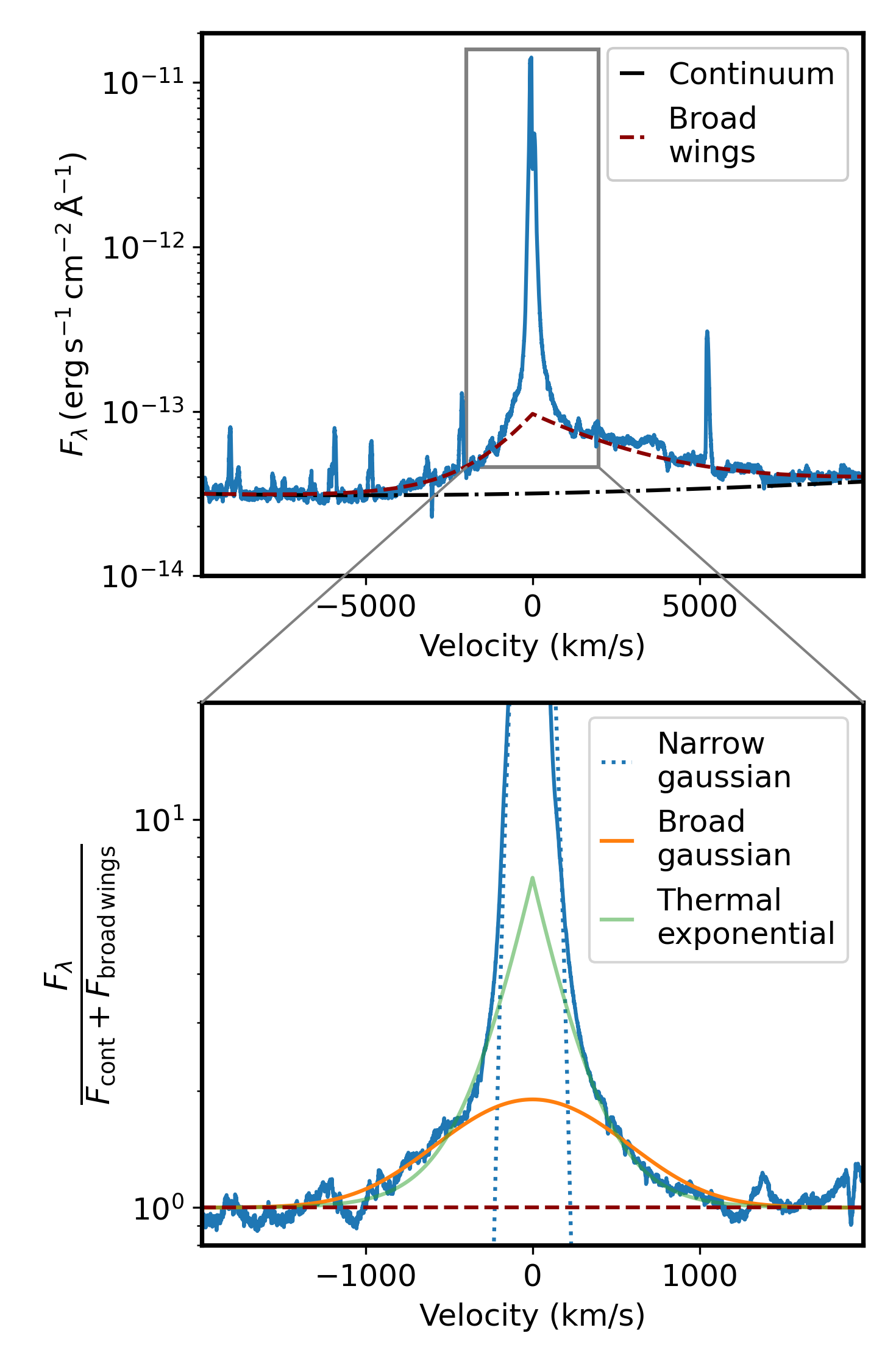}
    \caption{V1309\,Sco $H\alpha$ line-profile showing a broad-scale wing component (upper panel) and a zoom-in on the intermediate velocity component (lower panel). An exponential line-shape using a single thermal width (green) provides a superior match to the line-shape than a Gaussian (orange). The sharply decaying wings, centred around the narrow line, favour this intermediate component as scattering-related. }
    \label{fig:velocity_components}
\end{figure}

\subsection{Two-component electron-scattering: Upstream and downstream recombination}\label{sec:two_comp_es}
Between the regimes of narrow ($\!\sim\!100 \,\mathrm{km/s}$) and extremely broad features ($1000-10\,000 \,\mathrm{km/s}$), an intermediate velocity structure with $v\!\sim\!500\,\mathrm{km/s}$ is observed in the highest S/N object, V1309\,Sco, as shown in Fig.~\ref{fig:Electron_scattering_three_panel_upstream_downstream}. This could represent either the unscattered core of the downstream recombination emission or the photons produced upstream of the shock that travel inwards and bounce off the hotter ejecta. In contrast to previous regimes of electron-scattering modelling, here the narrow line region is outside the hot outflow. Whereas photons produced in the shocked ejecta must undergo many scatterings before escaping, photons coming upstream of the shock typically escape directly (producing a narrow line) or reflect once on inner ejecta (producing a thermally broadened shoulder); see Fig.~\ref{fig:shematic_electron_scattering}. 

In particular, this interpretation may be motivated on several accounts. 
First, the width of the intermediate component is around $500\,\mathrm{km/s}$, which is the thermal width of a single scattering. 
Second, the thermally broadened component shows tentative evidence for a blueward asymmetry. In the outflow geometry considered above, this suggests that the scattered photons are, on average, redirected into the observer’s line of sight by electrons moving towards the observer. In practice, this likely requires a significant change in photon direction upon scattering, as would occur if recombination photons are reflected or backscattered by a deeper layer. The effect is enhanced in proportion to how much faster the scattering layer expands than the layer emitting the recombination photons. 
Third, the relative luminosity of the (single scattering) thermal shoulder to the narrow emission is given by the product of the probabilities of a photon initially travelling inwards ($p_{inward}\!\sim\!0.5$), the probability that a photon interacts at the surface where $\tau_{es}=2/3$ ($p_{interaction}\!\sim\!0.5$) and scattering backwards towards a distant observer ($p_{backward,scattering}\!\sim\!0.5$). This suggests an order-of-magnitude difference between the luminosity of the narrow and the single-scattering broad component, which is indeed close to the observed ratio. 
Lastly, this intermediate component evolves in time similarly to the very narrow component and differently from the very broad component (as discussed in Sect.~\ref{sec:luminosity_es}).

In general, more complicated geometric structures (radial or angular) can provide explanations for why most photons always escape in the narrow component even with $\tau_{es}\gg1$. Indeed, equatorial-to-polar anisotropies are expected from the pre-merger ejection mechanism, which we explore in Appendix~\ref{app:equatorial_scattering}, and should motivate further development of electron-scattering codes in more general cases than the spherical symmetry limiting previous studies \citep[e.g.][]{Auer1972,Huang2018}. More complex geometries have also recently been invoked in light-curve modelling with 2D radiation-hydrodynamic simulations, which show that interaction between merger ejecta and equatorially concentrated circumbinary material can produce a blue peak followed by a redder plateau \citep{Kirilov2025}.

\begin{figure}
    \includegraphics[width=\linewidth]{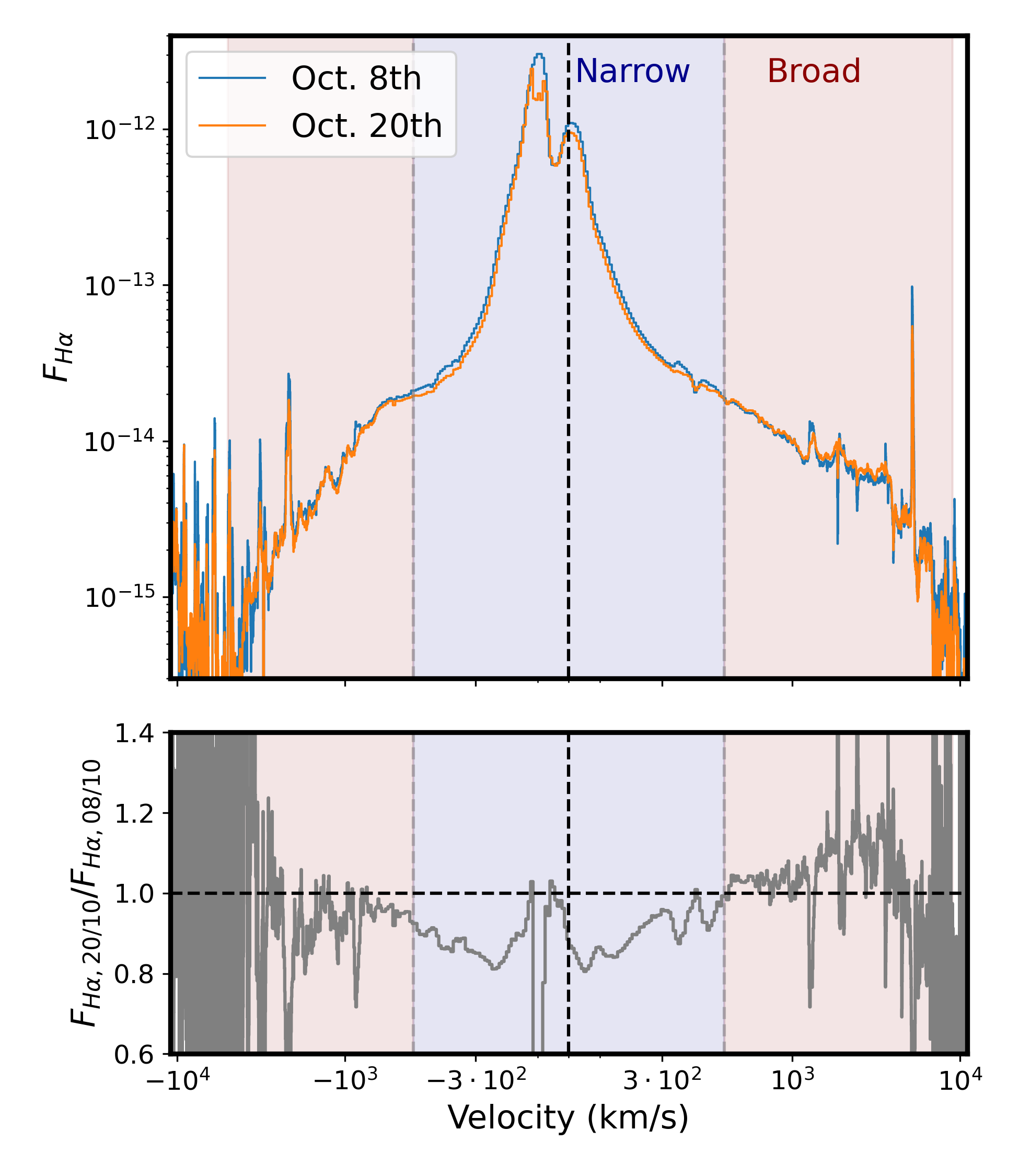}
    \caption{ Upper panel: $H\alpha$ line-profile for the two post-plateau epoch spectra of V1309\,Sco with the ratio indicated on the lower panel. Notably, the broad component increases relative to the narrow component with time. The intermediate velocity component (e.g. single scattering, $v\lesssim500 \,\mathrm{km/s}$) follows the narrow component, supporting the idea that these recombination photons share a similar physical origin. }
    \label{fig:temporal_broad_lines}
\end{figure}

The intermediate-velocity component could also contribute to the unscattered recombination line of the shocked ejecta, which is likely to emanate from a region with a typical expansion of several hundred kilometres per second. However, this interpretation -- where narrow and intermediate velocity components originate in different velocity-material -- does not naturally explain why the line-centroid of the narrow and intermediate components are consistent to within $\lesssim100$ km/s or why the time evolution of the intermediate-velocity component more closely matches the narrow component (see Fig.~\ref{fig:temporal_broad_lines} and Sect.~\ref{sec:luminosity_es}). Furthermore, as seen in Fig.~\ref{fig:velocity_components} from a goodness-of-fit perspective, parametrising the intermediate velocity component with an independent Gaussian profile cannot simultaneously match the broadest parts of the line at $500$-$1000$ km/s and the inner line at $200$-$500$ km/s (the narrow line dominates the flux below $\lesssim200$ km/s). Functionally, what is needed to match the data is something with sharply decaying wings centred on the narrow line, i.e. something analogous to the proposed reflective electron-scattering component.

\begin{figure*}
    \includegraphics[width=\linewidth]{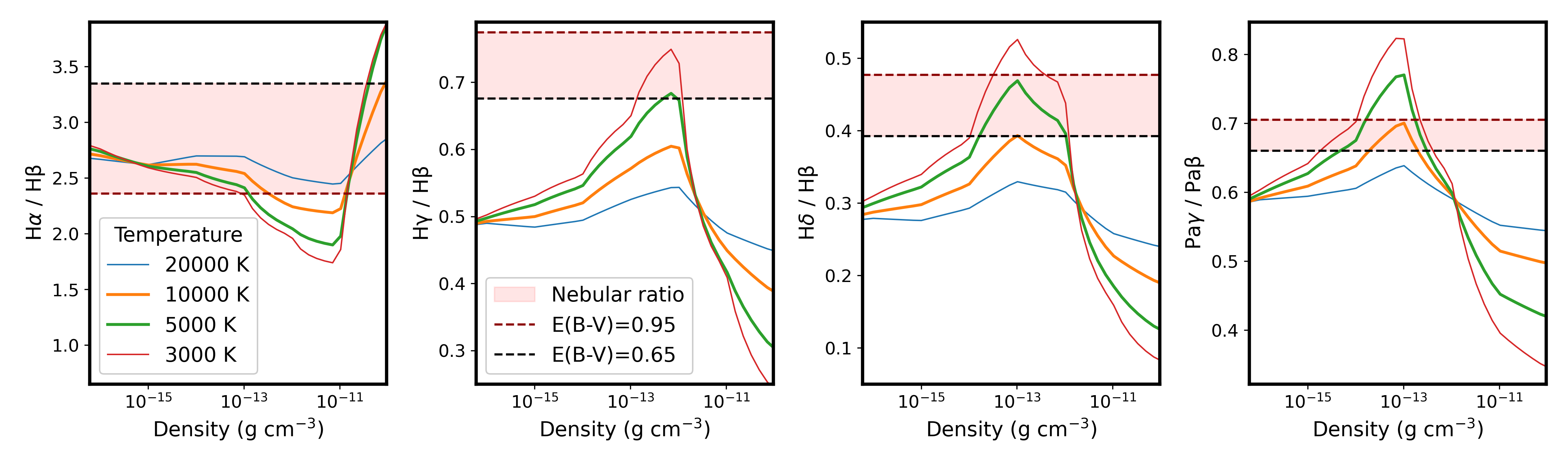}
    \caption{ Selected recombination line ratios relative as a function of density for various temperatures T=3000 K (red), T=5000 K (green), T=10\,000 K (orange), and T=20.000 K (blue). The lines show PyNeb calculations consistent with \cite{Ferguson1997}, while the observed V1309\,Sco line ratios are shown assuming E(B-V)=0.65 (dashed black line) and E(B-V)=0.95 (dashed red line). The various line ratios are only internally consistent in a limited range of parameters, i.e. $\rho_{\rm CSM}\!\sim\!10^{-12}-10^{-13} \mathrm{g/cm^3}$, $T\!\sim\!5000-10\,000 \mathrm{K}$, and near the observationally motivated range in dust extinction $E(B-V)=0.65-0.95$. }
    \label{fig:recombination_ratio}
\end{figure*}

\subsection{Luminosity from shock energetics}\label{sec:luminosity_es}
The luminosity of the forward shock can be estimated as
\begin{equation}
    L_{\rm sh} \!\sim\! 2\pi R_{\rm sh}^2 \rho_{\rm CSM} v_{\rm rel}^3,
\end{equation}
where $v_{\rm rel}$ is the shock velocity relative to the upstream CSM. For an order-of-magnitude estimate, we took the CSM to be slowly moving and adopted $R_{\rm sh}\!\sim\! v_{\rm sh}t$, with $v_{\rm sh}$ interpreted as a characteristic shock velocity at the interaction. This gives
\begin{align}
    L_{\rm sh} &\!\sim\! 2\pi \rho_{\rm CSM} v_{\rm sh}^5 t^2 \nonumber \\
    &\!\sim\! 10^{39} \,{\rm erg\,s^{-1}}
    \left(\frac{\rho_{\rm CSM}}{10^{-12} \,{\rm g/cm^{3}}}\right)
    \left(\frac{v_{\rm sh}}{300 \,{\rm km/s}}\right)^5
    \left(\frac{t}{30 \,{\rm days}}\right)^2 .
    \label{eq:plateau_interaction}
\end{align}

Shock deceleration could modify the normalisation at the order-unity level. We considered only the forward shock here as the reverse shock luminosity is expected to be smaller, although we consider the reverse shock in more detail in Paper~II. 
Regardless, even at low densities (motivated by the upstream recombination lines in Sect.~\ref{sec:Density}), the velocity inferred from the photospheric expansion (or equivalently from the electron-scattering wings) implies that the shock power is sufficient to produce the entire luminosity of these transients. A wide range of luminosities might naturally be expected due to the sensitive dependence on $v_{sh}$ in particular.

We next evaluated the luminosity of the hot upstream ionised component itself to check whether this should be detectable. Given the emissivity, $\epsilon = n_e^2 \Lambda$ and the volume, $V=4\pi R_{sh}^2 \Delta R = 4\pi R_{sh}^3 \frac{t_{cool}}{t}$, the associated luminosity is $\epsilon V$. Here, $\Delta R = R_{sh}\cdot t_{cool}/t$ represents the length scale over which cooling takes place, with the fractional scale $\Delta R/R_{sh}$ given by the ratio of the cooling time, $t_{cool}$, to the total time post-merger, $t$).  We then obtain
\begin{equation}
    L_{\mathrm{electrons,u}} = (n_e^2 \Lambda) \left(4 \pi R_{sh}^3 \frac{t_{cool}}{t}\right) = 4\pi R_{sh}^3n_e\frac{k_B T}{t} \!\sim\! \frac{k_B T N_{e,tot}}{t},
\end{equation}
where the cooling timescale is $t_{cool}=\frac{k_B T}{\Lambda n_e}$ and the cooling function is $\Lambda\gtrsim10^{-26} \,\mathrm{erg \,cm^{3}\,s^{-1}}$ \citep{Schure2009}. The upstream densities ($\rho_{CSM}\!\sim\!10^{-12} \frac{g}{cm^3}$ derived from the line ratios; see Sect.~\ref{sec:Density}) are sufficient to ensure cooling much faster than the dynamical timescale ($t_{cool}/t\ll1$). The ionised upstream should thus be geometrically thin, which emphasises that the electron-scattering region should be located in the denser downstream, where the optically thick environment hinders efficient cooling and substantial outflow velocities can produce the electron-scattering wing asymmetry. 
Furthermore, the thinness of the upstream ionised layer implies a modest upstream electron-number, $N_{e,tot}\!\sim\!10^{57}$, and therefore a weak relative luminosity: 
\begin{equation}
    L_{\mathrm{electrons,u}}\!\sim\! 10^{36} \mathrm{\frac{erg}{s}} \left(\frac{T}{10\,000 K}\right) \left(\frac{\rho_{CSM}}{10^{-12} \mathrm{g/cm^3}}\right).
\end{equation}

The shock energy is likely to enter the system as UV or soft X-ray photons, which photoionise hydrogen in the surrounding region and subsequently produce recombination emission. Recombination lines propagating into the hot-electron region undergo electron-scattering, while in the outer envelope of the ejecta lines are partially reprocessed by H$^{-}$-opacity towards a blackbody continuum for all photospheric epochs. The photosphere then expands at roughly constant luminosity until it reaches a radius at which the effective temperature falls below that required for efficient H$^{-}$-opacity (e.g. $T=L/(4\pi r^2\sigma)\lesssim 3500\,$K). At that point, the recombining region upstream of the shock becomes directly visible, revealing a non-thermal continuum, stronger recombination lines, and free-bound emission. We note, however, that in sufficiently dense upstream environments the gas may remain optically thick even after H$^{-}$ opacity becomes inefficient.

Furthermore, because the shock is energetic enough to ionise the gas as it passes through it, cooling in the shocked ejecta should produce an additional recombination luminosity. Specifically, each swept-up particle can in principle radiate of order one hydrogen ionisation energy, $\chi_H$, upon recombination, so that sweeping up mass at a rate $\dot{M}_{\rm swept}$ gives a downstream recombination luminosity
\begin{equation}
    L_{H,\rm downstream} \!\sim\! \dot{M}_{\rm swept}\frac{\chi_H}{m_p}
    = 4\pi R_{\rm sh}^2 v_{\rm sh}\rho_{\rm CSM}\frac{\chi_H}{m_p},
\end{equation}
where $R_{\rm sh}$, $v_{\rm sh}$, and $\rho_{\rm CSM}(R_{\rm sh})$ are all functions of time. As discussed in Sect.~\ref{sec:two_comp_es}, if the downstream recombination line photons interact with a population of Thomson-thick electrons, the lines can be scattering-broadened to produce the observed very broad component. This expression should not be interpreted as an upper limit on the total recombination luminosity. A gas parcel may undergo multiple ionisation-recombination cycles if it is irradiated by ionising photons from the shock or cooling layer. Thus, $L_{H,\rm downstream}$ is a conservative estimate of the broad component's luminosity, as the shocked and unshocked ejecta (if sufficiently ionised) may also contribute additional recombination luminosity. 

The key conceptual point is that recombination upstream and downstream of the shock need not evolve in the same way. Upstream of the shock, the recombination luminosity is powered by photoionisation from the shock radiation field and should therefore scale with the shock luminosity, $L_{H,\rm upstream} \propto L_{\rm sh} \propto R_{\rm sh}^2 \rho_{\rm CSM} v_{\rm sh}^3$. By contrast, downstream recombination instead scales with the rate at which material is swept up, $L_{H,\rm downstream} \propto R_{\rm sh}^2 \rho_{\rm CSM} v_{\rm sh}$. Thus,
\begin{equation}
    L_{H,\rm upstream} \propto v_{\rm sh}^2 L_{H,\rm downstream}.
\end{equation}
As long as the shock decelerates, the upstream-powered narrow component is therefore expected to fade faster than the downstream-powered broad component. Thus, the narrow and broad components are distinguished not only by velocity but also by their temporal evolution.

Indeed, Fig.~\ref{fig:temporal_broad_lines} shows the temporal evolution of H$\alpha$ in V1309\,Sco, the only object in our sample with clear exponential wings observed over multiple epochs. The narrow component, together with the single-scattering thermal-width component, fades relative to the broad component. This both corroborates the distinct physical origin of the narrow and broad photons and supports the interpretation of the single-scattering component as arising from upstream recombination. We note, however, that the situation near spectral peak is somewhat complicated by possible self-absorption in the line and by saturation in the later spectrum, as indicated by the flat-topped line peak. The temporal dependencies of the upstream and downstream luminosities are directly predictable given the ejecta and CSM structure, which can be compared against the predictions on the light curve evolution of LRN (evolution of photospheric radius and temperature), which we will cover in Paper~II.

\begin{figure*}
    \includegraphics[width=\linewidth,viewport=10 10 980 502 ,clip=]{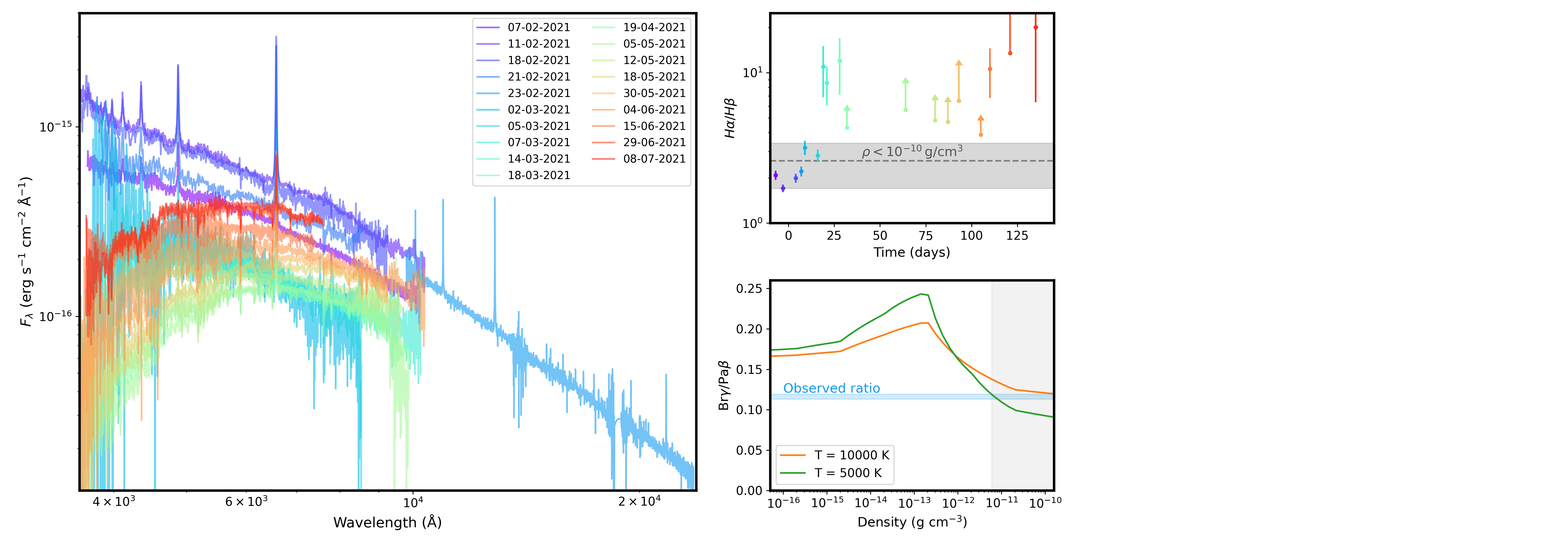}
    \caption{Spectra of the LRN AT2021blu \citep{Pastorello2023} with subplots highlighting (top right) the Balmer decrement as a function of time and (bottom right) the ratio of NIR hydrogen lines.
    The observed NIR line-ratio is inconsistent with the density from V1309\,Sco, with instead a higher implied density of $\rho_{\rm CSM}\gtrsim10^{-11} \,\mathrm{g \,cm^{-3}}$.
    The relative weakening of H$\beta$ is unlikely to be due to dust-formation, as this scenario would require a rapid formation without corresponding reddening of the spectral continuum.
    The intrinsic recombination ratio may be suppressed by collisional de-excitation of n=4 for $\!>\!10^{-10} \mathrm{g \,cm^{-3}}$ or due to radiation transport effects between the recombination region and the photosphere, where photons in H$\alpha$ and H$\beta$ can have different escape probabilities. Regardless, both these probes suggest significantly denser CSM environments in AT2021blu than V1309\,Sco. Under a shock interpretation, this order-of-magnitude CSM higher density (and the slightly larger velocity from blackbody fits to photometry) requires AT2021blu to be a hundred-fold more luminous than V1309\,Sco, as observed. }
    \label{fig:AT2021blu_spectra}
\end{figure*}

\section{Density from recombination line ratios}\label{sec:Density}

A notable prediction of the proposed interpretation of LRNe is that collisions with a surrounding low-density medium are sufficient to provide the observed luminosities. Furthermore, the broad range of luminosities in LRNe should be a consequence of the variations in characteristic velocity of the ejecta and in the CSM density. In the following, we discuss testing these predicted low densities against the values of $\rho_{\rm CSM}$ inferred from narrow hydrogen recombination-line ratios (see Figs.~\ref{fig:recombination_ratio} and~\ref{fig:AT2021blu_spectra}). In particular, whereas a recombination-powered interpretation of LRNe demands very large ejecta masses \citep[up to several hundred solar masses; e.g.][]{Matsumoto2022}, the masses required in the shock-powered scenario are far more modest, with $\rho_{CSM}\!\sim\! 10^{-13}-10^{-9} \mathrm{g/cm^3}$ (e.g. given the characteristic volumes, $M\!\sim\!10^{-4}-0.1\,M_{\odot}$) for $L\!\sim\!10^{38}-10^{41} \,\mathrm{erg/s}$. In the case of the faintest LRN, such as V1309\,Sco, shock energetics with $L\!\sim\! 10^{38}-10^{39} \mathrm{erg/s}$ imply $\rho_{\rm CSM}\!\sim\! 10^{-13}-10^{-12} \,\mathrm{g/cm^3}$, while the brightest LRNe must be embedded in media that are one to a few orders of magnitude denser and/or have larger ejecta velocities. 

\subsection{Narrow hydrogen line ratios ($\rho_{CSM}$})

Figure~\ref{fig:recombination_ratio} illustrates the observed line ratios from the post-plateau epochs of V1309\,Sco, together with the line ratios expected as a function of density from PyNeb calculations, which assume Case~B recombination and neglect radiation-transport effects after the photons are emitted. Using post-plateau spectra is optimal, as it implies that there is no optically thick region between the recombination region and the observer. Notably, for the observed line ratios to be mutually consistent, a reddening of $E(B-V)\!\sim\!0.7-0.9$ is required, which is constrained primarily by the bluer recombination lines ($H\alpha/H\beta$, $H\gamma/H\beta$, etc.) and, though statistically independent, is broadly in agreement with estimates of Galactic reddening towards the source \citep[e.g. $E(B-V)=0.84\pm0.04$ from  K I equivalent width; see][]{Mason2022}. The temperature must be around 5,000-10,000\,K, because if it were much hotter than this, the strength of H$\gamma$, H$\delta$, Pa$\gamma$ would be suppressed. The density is particularly well-constrained from the NIR Paschen recombination lines, which are less sensitive to dust reddening. Remarkably, for V1309Sco the recombination-line inferred density $10^{-13}$-$10^{-12} \,\mathrm{g/cm^3}$ is close to that of the predicted shock interpretation. 

By contrast, LRNe embedded in denser CSM enter the regime where the higher hydrogen levels approach their critical densities, leading to a relative suppression of transitions from these states. Under standard Case~B recombination, one expects $H\beta/H\alpha \approx 0.3-0.4$, so the much smaller observed ratios indicate a substantial departure from this limit. Indeed, increasingly luminous LRNe appear to show large deviations from Case~B recombination. Post-plateau spectroscopy of V838 Mon and M101-2015OT1 (at $10^{39}-10^{40} \mathrm{\,erg /s}$), yields $H\beta/H\alpha=0.051\pm0.001$ and $H\beta/H\alpha\!<\!0.1$, respectively, showing that these more luminous LRNe have reached the critical-density regime  (see Appendix~\ref{sec:BB_fit}).

In Fig.~\ref{fig:AT2021blu_spectra}, we show the observed photospheric spectra of AT2021blu, the H$\beta$/H$\alpha$ ratio over time, and the Br$\gamma$/Pa$\beta$ ratio. As noted, photospheric epochs are less straightforward to interpret because of potential reprocessing \citep{Netzer1975,Krolik1978ApJS,Drake1980ApJS}, but several interesting constraints can still be extracted.
First, the relative weakness of transitions from higher $n$ (e.g. Br$\gamma$/Pa$\beta$ ratio, lower right panel) suggests we are in the regime where critical densities suppressing higher-order lines --- i.e. at $\gtrsim10^{-11} \,\mathrm{g/cm^{3}}$ beyond the densities inferred in V1309\,Sco. These near-infrared transitions are more interpretable because the continuum opacity is lower, dust reddening is insignificant, and the small Einstein A-coefficients imply these are less affected by line-trapping. 
Second, the ratio of H$\beta$/H$\alpha$ displays dramatic evolution in time beginning near Case~B values in the initial luminosity peak but rapidly drops between 10 and 20 days (upper right panel). Such a rapid drop is unlikely to be due to dust reddening, as i) it would require a rapid formation of dust from $A_V\lesssim0.06$ to $A_V\!\sim\!3$ between 10-20 days and ii) it occurs without a corresponding dramatic reddening of the continuum. Such strongly suppressed H$\beta$/H$\alpha$ ratios cannot be produced as intrinsic recombination ratios for $\!<\!10^{-10} \mathrm{g \,cm^{-3}}$ within the PyNeb calculations. One could invoke differential reprocessing or large self-absorption of H$\alpha$ and H$\beta$ as an explanation. Bound-free opacity can at most raise $\kappa_{abs}$ by a factor of several around $H_{\alpha}$ compared to $H_{\beta}$, leading to stronger reprocessing of $H_{\alpha}$ relative to H$\beta$. While H$\alpha$ photons are resonantly scattered and eventually escape as H$\alpha$, $H_{\beta}$ photons can frequently be converted into H$\alpha$ photons through cascades, so to attain the observed H$\beta$/H$\alpha$ ratios only provides modest constraints with column densities $N_H\gtrsim10^{23} \mathrm{cm^{-2}}$ \citep{Netzer1975}. Thus, the NIR ratios of higher-state lines remain a cleaner diagnostic. 
While shock energetics and recombination line ratios agree on the density, we leave detailed opacity considerations needed to produce the thermal reprocessing implied from the continuum to future work.

\subsection{Broad hydrogen line intensity ($T_{H\alpha}$})
The broad line forms in the shock's downstream (or in the ejecta) as required by the observed outflow velocity of several hundred kilometres per second. The dense downstream is optically thick in the Balmer lines even when the system is otherwise in a nebular phase, so line luminosities and line ratios should be interpreted with caution. In the optically thick limit the emergent line intensity is capped by the local source function, which in LTE approaches a blackbody, $S_\lambda\!\sim\!eq B_\lambda(T)$. We therefore write the integrated broad-line luminosity as
\begin{equation}
    L_{H\alpha}=\int L_{\lambda}\,d\lambda \approx 4\pi R_{sh}^2 \pi B_{\lambda}(T_{H\alpha}) \,\Delta\lambda,
\end{equation}
where $B_{\lambda}(T_{H\alpha})$ is the local blackbody evaluated in the $H\alpha$ line-forming region and $\Delta\lambda$ is the line-forming width, which is dominated by the expansion velocity when discussing the optically thick line-centre (e.g. $\Delta\lambda\approx\lambda_0 v_{sh}/c$). Importantly, the optically thick Balmer lines imply that line luminosity is not a direct probe of the ionised mass, but a useful diagnostic of the local radiative transfer conditions. Matching the broad-line H$\alpha$-intensity in both faint and bright LRNe (e.g. V1309\,Sco and NGC4490-2011OT1, respectively) suggests similar downstream conditions across objects with $T_{H\alpha}\!\sim\!5000$-$10\,000\,$K given $R_{sh}$/$v_{sh}$ from light-curve fitting (see Appendix~\ref{sec:BB_fit}). This is consistent with the gas temperature inferred from electron-scattering (see Sect.~\ref{sec:electron_constraints}). This temperature range can accommodate the factor of a few in broad-line ratio observed in H$\alpha$/H$\beta$ for V1309\,Sco, although this would also be broadly consistent with Case~B recombination ratios.

\begin{figure}
    \includegraphics[width=\linewidth]{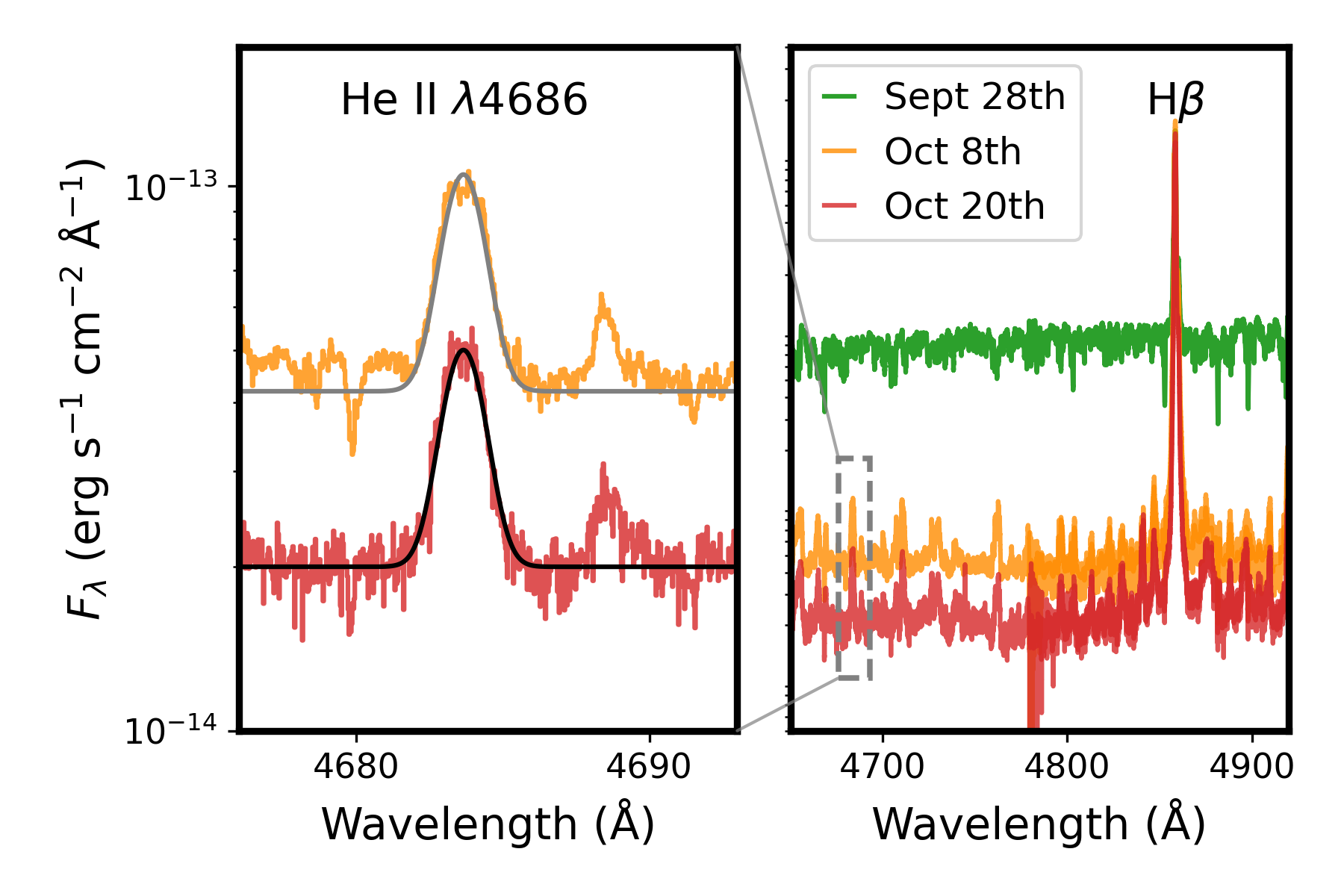}
    \caption{High-ionisation He\,II $\lambda 4686$ emission in V1309\,Sco. The He\,II line fades coherently with the continuum and with other recombination lines (see right panel), indicating a common energetic origin. The detection requires a hot ionising component, with characteristic temperatures $T\gtrsim5\times10^{4}$\,K. At the same time, the low luminosity ratio $L_{\rm He\,II\,\lambda4686}/L_{\rm H\beta}\!\sim\!0.003$ lies well below the predictions of unsuppressed fast radiative-shock models. This suggests either relatively low shock velocities and/or suppression of the hot ionising luminosity by rapid downstream cooling and turbulent mixing. }
    \label{fig:HeII4686}
\end{figure}

%\subsection{Luminosity d}
\subsection{High-ionisation lines}\label{sec:he4686}
A potential issue for shock-powered models is the weakness, or absence, of high-ionisation recombination lines, in particular He\,II $\lambda 4686$ (Fig.~\ref{fig:HeII4686}). Ionising to He\,III requires photons with $h\nu\!>\!54.4$\,eV; this is not implausible for the immediate post-shock gas and is predicted by the blue nebular component in Fig.~\ref{fig:mstar}. The challenge is how substantial a He\,III emitting region can be maintained. The temperatures, $T_e\!\sim\!5000-10000\,{\rm K}$, inferred for the bulk observed ionised gas whether from electron-scattering wings or hydrogen recombination-line ratios, are sufficient to keep hydrogen ionised but are not sufficient for He\,III ionisation. The weak He\,II $\lambda 4686$ emission in the post-plateau phase from the high S/N spectra of V1309\,Sco \citep{Mason2022}, together with its non-detection in other LRNe \citep{Cai2022}, should therefore be interpreted as a constraint on the ionisation structure and cooling efficiency of the shock-powered emission, rather than as direct evidence against shocks. 
In 1D radiative-shock calculations the strength of He\,II recombination emission is a steep function of shock velocity and of the ionisation structure of the shock and its precursor. For example, the MAPPINGS~III models predict He\,II $\lambda4686/{\rm H}\beta\simeq0.05$ at $v_{\rm sh}=200\,{\rm km\,s^{-1}}$, rising to $\sim0.3$ for $v_{\rm sh}\gtrsim500$--$1000\,{\rm km\,s^{-1}}$ \citep{Allen2008}. Multi-dimensional radiative shocks are even less efficient at maintaining a large volume of He\,III gas, because thin-shell instabilities and turbulent mixing with cold material can reduce the luminosity emerging at temperatures comparable to the immediate post-shock temperature by more than an order of magnitude \citep{Steinberg2018}. Thus, for relatively low shock velocities and a stratified cooling layer shock emission can produce comparable line-ratios to the observed He\,II $\lambda4686/{\rm H}\beta\sim0.003$ in V1309\,Sco.
 
Conversely, the detection of even weak He\,II $\lambda 4686$ emission implies that some hot ionising component is present with characteristic temperatures $\gtrsim5\times10^{4}$\,K. The plethora of coherently fading recombination lines in V1309\,Sco (including H\,I, Fe\,II, He\,I, and He\,II; see Fig.~\ref{fig:HeII4686}), thus highlights a stratified ionised medium. A full analysis of these various elements line-ratios could further inform the shock velocity and elucidate the ionisation structure.

\section{Discussion and conclusions}\label{sec:discussion}

Having introduced several new observational avenues for constraining the physics of LRNe (i.e. electron-scattering wings, temperature stratification from weak high-ionisation lines, and density constraints from recombination line ratios), we briefly summarise the main findings.

\begin{enumerate}    
    \item Two distinct spectral components are found in post-plateau epochs: i) a blue recombination spectrum indicating a hot environment and ii) a cold stellar-like continuum. In photospheric epochs the former is partially reprocessed, while these two fade coherently over time for the two objects with multiple post-plateau spectra. We propose that they represent the upstream and downstream shells illuminated by a common shock surface.

    \item We identify two electron-scattering components: very broad red-skewed wings ($v\!\sim\!1000-10\,000\,\mathrm{km/s}$) from multiply scattered downstream photons and an intermediate-width component ($v\!\sim\!500\,\mathrm{km/s}$), plausibly produced by single scattering of upstream recombination photons. The inferred hot outflow has a characteristic temperature $T_e\!\sim\!10\,000 \,\mathrm{K}$ and outflow-velocity $v_{es}\!\sim\!400\,\mathrm{km/s}$. 

    \item Low densities, $\rho_{\rm CSM}\!\sim\!10^{-13}$-$10^{-9} \mathrm{\,g/cm^{3}}$, and low velocities, $\lesssim100\,{\rm km/s}$, are implied for the upstream ambient medium by hydrogen recombination lines in the post-plateau phase. The brighter LRNe show deviations from Case~B recombination suggesting higher densities. Combined with the velocities inferred from electron-scattering, the expanding blackbody, or the escape speeds of the binary progenitors, these densities are sufficient for shock interaction to dominate the energetics. The diversity of luminosity may here be the predictable outcome of the observed diversity of CSM densities (and ejecta velocities).  

    \item To explain simultaneously the energetics, the slow upstream velocities above a rapidly expanding photosphere, and the coexistence of a cold blackbody, hot electrons, recombination lines, and weak high-ionisation lines, we argue that these transients are powered by shock interaction. Recombination lines are expected to be produced in the immediate upstream CSM, which is photoionised by UV or soft X-ray photons produced at the forward shock; in the outermost layers of the unshocked ejecta, which may be likewise photoionised by the reverse shock; and in the fast-cooling shocked CSM and ejecta, which are ionised directly by the forward and reverse shock, respectively. These recombination lines, together with the continuum generated by shock interaction, are themselves largely reprocessed by H$^-$ ahead of the shock (see Fig.~\ref{fig:shematic_electron_scattering}), initially producing a $\!\sim\! 5000$\,K blackbody continuum. As the shock expands, the shell of ionisation moves outwards along with it, similarly to the photospheric surface. During the photospheric phase, the blackbody surface remains outside of the shock, so that narrow recombination lines with $v_{rec}\lesssim100\,\mathrm{km/s}$ and a hot outflowing blackbody with $v_{es}\!\sim\!400\,\mathrm{km/s}$ are simultaneously observed. The transition to the post-plateau phase occurs once the photosphere recedes inside the shock, revealing higher-velocity ionised regions. 
    
\end{enumerate}
In a companion paper, we explore how LRN light curves can be explained by shock breakout and continued interaction (Paper~II).

\begin{figure}
    \includegraphics[width=\linewidth]{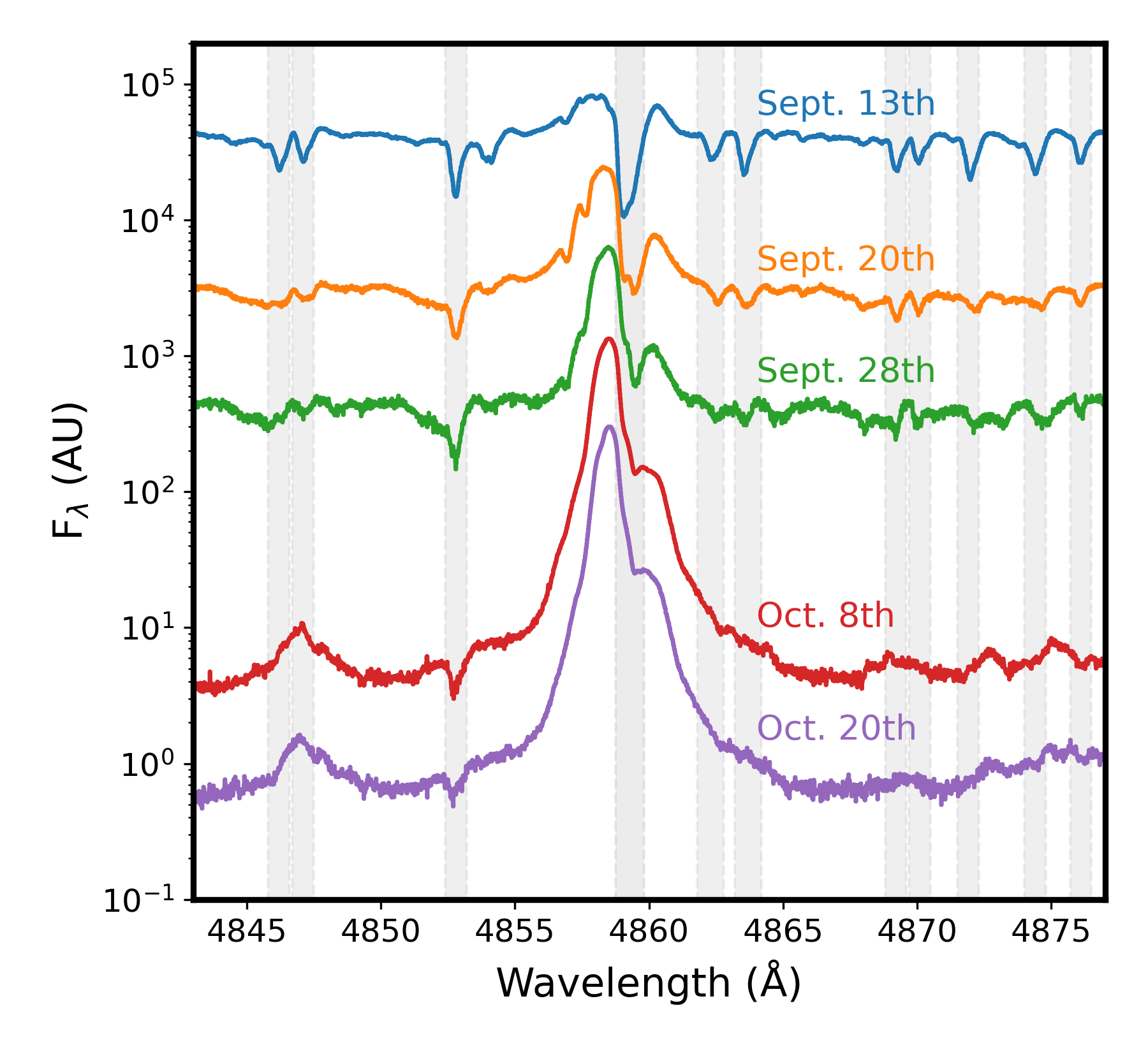}
    \caption{Spectra of V1309\,Sco around the H$\beta$ line in the photospheric (blue, orange, and green) and the post-plateau epochs (red and purple). A series of small-scale velocity features with $v\lesssim100\,\mathrm{km/s}$ is observed (highlighted with grey shading) due to slowly expanding pre-merger CSM material surrounding the more rapidly expanding photospheric surface. These features weaken across photospheric epochs as the surrounding CSM is increasingly swept up. 
    }
    \label{fig:disappearance}
\end{figure}

\subsection{Shocks as the energetic sources of LRN}
The energetic sources of LRNe have remained a matter of much debate. However, the observations now require three elements of a shock interpretation.

First, we note a photospheric surface of fast ejecta moving into a slower ambient medium. This slower surrounding CSM is characterised by a decreasing column density of material, as indicated by the weakening of upstream narrow absorption features (see Fig.~\ref{fig:disappearance}). This is in contrast to the plateau-phase in type II-P SNe, where absorption in hydrogen lines typically increases with time as the photosphere recedes inwards in mass coordinate \citep[e.g.][]{Gutierrez2017}. In these systems, absorption components can grow very deep -- even extending below the continuum.

Second, the density from recombination line ratios in a faint LRN such as V1309\,Sco requires a low ambient medium density, $\rho_{\rm CSM}\!\sim\!10^{-13}-10^{-12} \mathrm{\,g/cm^{3}}$, which in combination with the hot component's velocity suggests that the shock luminosity is sufficient to power the event. For the correspondingly more luminous LRN, such as AT2021blu, the weak $Br\gamma/Pa\beta$ and large changes in $H\beta/H\alpha$ motivates a higher density. While the recombination regions do not need to represent the ejecta as a whole, none of the densities inferred from line ratios imply the high masses, $O(1-100M_{\odot})$, needed when hydrogen recombination to power LRNe \citep{Matsumoto2022}.

Third, the asymmetric electron-scattering wings suggest that we are now explicitly seeing a hot ejecta-component undergoing a radial outflow. Recombination photons appear to originate both upstream and downstream of this hot component. The outflowing nature (as well as the characteristic velocity being similar to the ejecta) suggests that this component does not arise from a central accretion-powered source \citep[e.g.][]{Soker2020}.

\subsection{Predictions and future work}
This work opens several avenues worthy of further exploration. These include i) interaction-produced light curves for slow shocks (as we discuss in Paper~II); ii) electron-scattering wings beyond the simple spherical hot outer regions relevant in previous regimes; iii) modelling of high ionisation lines and the potentially turbulent stratification of the downstream; iv) extensions of these analyses to late time-spectra of 'intermediate luminosity red transients', which display similarly broad lines \citep{Valerin2024}; v) the time-scale in post-plateau epochs for various spectral signatures to decay (e.g. electron-scattering wings and recombination-lines); and vi) detailed comparisons to the opacity predictions in this regime. The last point includes elements such as examining how the photospheric radii should evolve in time for the relevant opacity in each epoch but also readily testable predictions on the temporal evolution of recombination lines. In particular, the relative reprocessing opacities can be seen when investigating recombination lines, in particular the H$\alpha$ luminosity and the Balmer decrement in time (i.e. $H_\alpha/H_\beta$). 
While we demonstrated that modest densities can account for the overall energetics of LRNe, an outstanding challenge is to explain how the upstream material provides sufficient opacity to reprocess the emission into the observed thermal blackbody continuum of the photospheric phase. Resolving this regime of red and relatively cold radiative transfer will likely have broader implications, such as explaining the high-redshift 'little red dot' \citep[LRD;][]{Matthee2024} population with their phenomenologically similar unexplained blackbodies \citep{deGraaff2025b}, cold-gas envelopes, and strong electron-scattering systems \citep{Rusakov2025,Sneppen2026,Wang2026}.

Ultimately, detailed comparisons are needed to validate the robustness of the trends predicted. We hope this motivates future observations that specifically target the time around $t_{pl}$, when the blackbody temperature approaches 3000-3500\,K. At this point, the likely cessation of H$^-$-opacity precipitates a dramatic evolution of both the continuum and the recombination-lines, revealing the inner ejecta's energetics to the outside universe. 
Furthermore, the complete absence of spectra covering the precursor rise to shock breakout points to an additional and largely unexplored regime for constraining the radiative properties of LRNe. Given the plethora of potential detections within optical/NIR time-domain astronomy from upcoming surveys, including the recent introduction of the \textit{Vera C. Rubin Observatory}, such rising precursors, on timescales of several months, may provide valuable tests.

\begin{acknowledgements}
The authors would like to thank Ehud Nakar, Dan Kasen, Darach Watson and Jim Fuller for helpful discussions on interpreting these systems. Additionally, we express our gratitude to Andrea Pastorello, Elena Mason and Ulisse Munari for sharing data and thoughts on reduction-pipelines. The Cosmic Dawn Center (DAWN) is funded by the Danish National Research Foundation under grant No.~140. AS is funded in part by the European Union (ERC, HEAVYMETAL, 101071865). Views and opinions expressed are, however, those of the authors only and do not necessarily reflect those of the European Union or the European Research Council. Neither the European Union nor the granting authority can be held responsible for them.  KH and CMI are supported by the JST FOREST Program (JPMJFR2136) and the JSPS Grant-in-Aid for Scientific Research (20H05639, 20H00158, 23H01169, 23H04900).
\end{acknowledgements}

\bibliography{refs}
\bibliographystyle{aa}

\appendix
\section{Monte Carlo model for electron-scattering wings}

\begin{figure*}[b]
    \centering
    \includegraphics[width=\linewidth]{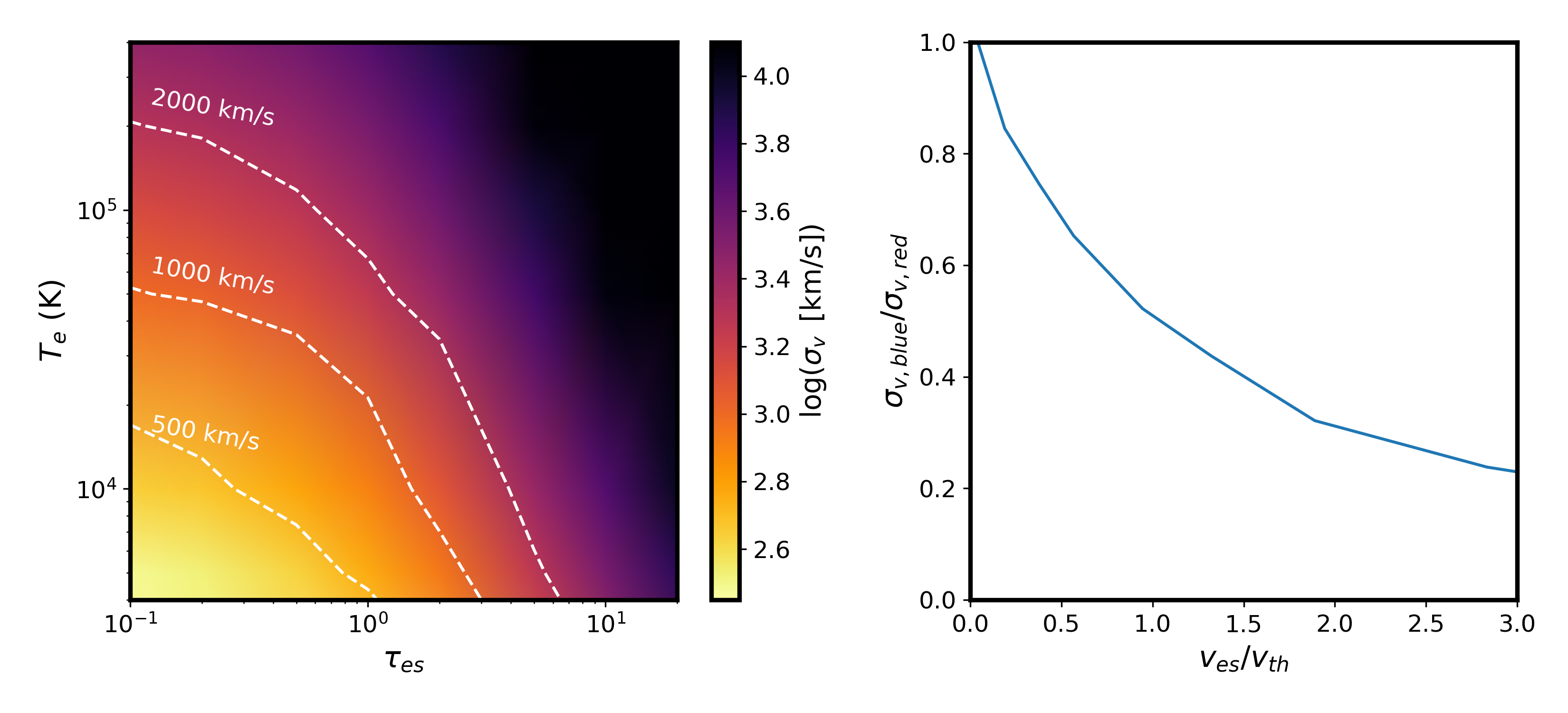}
    \caption{E-folding velocity of exponential electron-scattering wings as a function of the key parameters $T_e$, $\tau_{es}$, and $v_{es}$. 
    Left panel: overall e-folding scale across the optical-depth–temperature plane. 
    Right panel:degree of asymmetry as a function of the outflow velocity, $v_{es}$, expressed in units of the thermal velocity, 
    $v_{th}=\sqrt{2k_BT/m_e}$, i.e. the characteristic thermal velocity per scattering.}
    \label{fig:wings_mc}
\end{figure*}

\begin{figure}[t]
    \includegraphics[width=\linewidth]{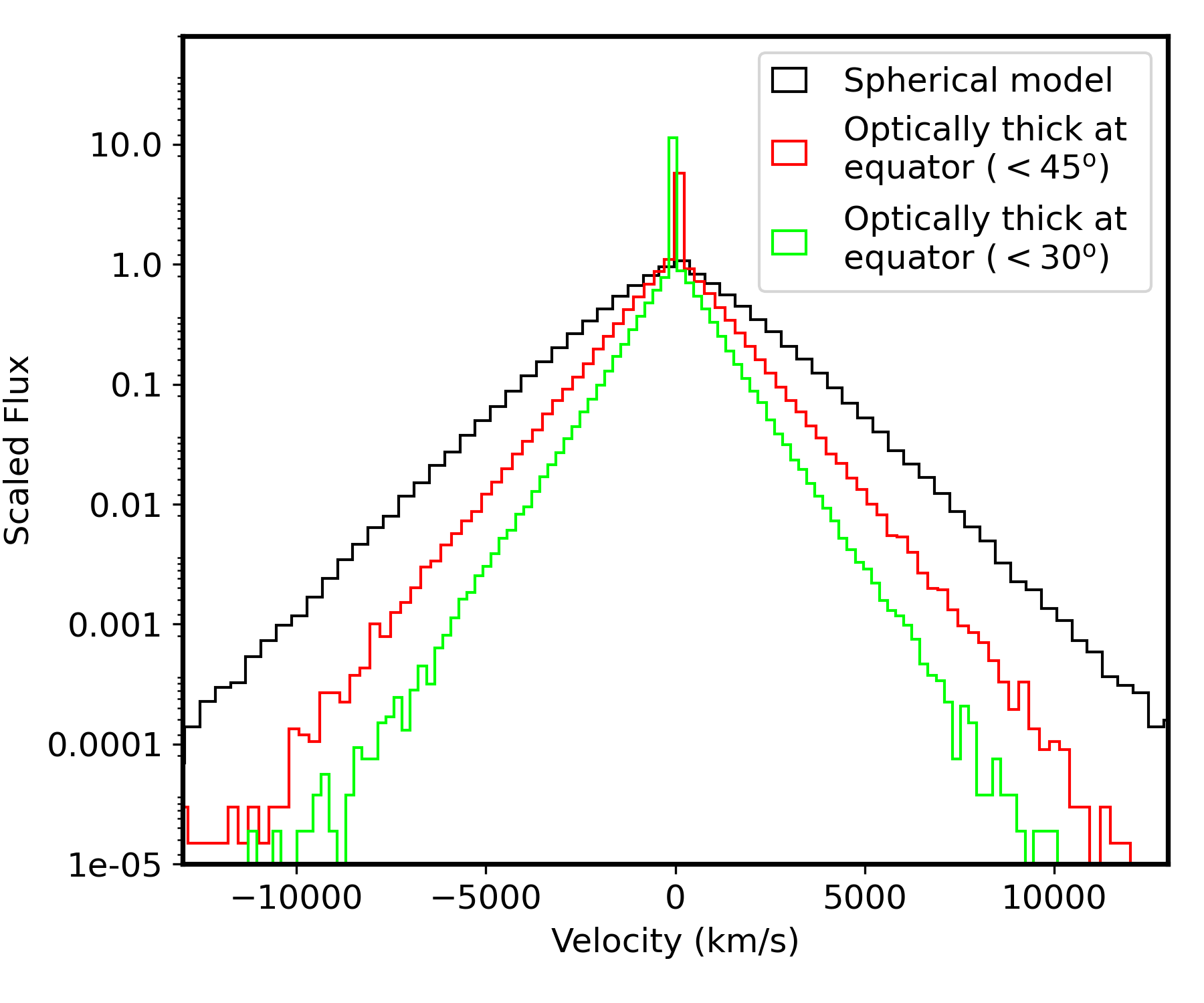}
    \caption{Electron-scattering wings for similar parameters ($\tau_{es}=5,T=10\,000\,K,v_{es}=0$), but with three different geometries, i) a spherical hot electron region, ii) an equatorial hot electron region (with an optically thick region confined to $\!<\!45^\mathrm{o}$ from the equatorial plane) iii) a more confined equatorial hot electron region (with an optically thick region confined to $\!<\!30^\mathrm{o}$ from the equatorial plane).}
    \label{fig:equatorial}
\end{figure}

To investigate how the electron-scattering wings depend on the key parameters $T_e$, $\tau_{es}$, and $v_{es}$, we constructed a Monte Carlo simulation of photon diffusion from a central region \citep[e.g.][]{Huang2018}. Photons are injected at $r=0$ and propagated until they escape the scattering region $r\!>\!R_{outer}$. Figure~\ref{fig:wings_mc} shows the resulting e-folding velocity scale of the electron-scattering wings, $\sigma_v$, as a function of the model parameters. In the left panel, $\sigma_v$ is shown across the $T_e$--$\tau_{es}$ plane. Although different combinations of temperature and optical depth can produce similar e-folding scales, two regimes of particular interest both require $\tau_{es}\gg1$: (i) the regime of broad lines with efficient cooling, corresponding to $T_e\!\sim\!10^4\,\mathrm{K}$ in the left panel, and (ii) the regime in which the thermal and bulk outflow velocities are comparable, $v_e/v_{th}\!\sim\!1$, shown in the right panel.

In Fig.~\ref{fig:wings_mc} (right panel), we show the blue-to-red wing asymmetry as a function of the outflow velocity, expressed in units of the thermal velocity of the scattering electrons, $v_{th}=\sqrt{2k_BT/m_e}$. For reference, the exponential wings in V1309\,Sco have e-folding scales of $\sigma_{v,blue}\!\sim\!800\,\mathrm{km/s}$ and $\sigma_{v,red}\!\sim\!1600\,\mathrm{km/s}$. The observed factor-of-two asymmetry therefore suggests $v_{out}\!\sim\! v_{th}$.

\subsection{Electron-scattering wings in an aspherical geometry}\label{app:equatorial_scattering}
In Fig.~\ref{fig:equatorial}, we show the electron-scattering wings resulting from a Monte Carlo simulation of photon diffusion from a central region \citep[analogous to][]{Huang2018}, but where we examined various geometries. In particular, we compared a spherical model with models in which the hot, optically thick material is confined to equatorial regions, with no corresponding material in the polar ejecta. Photons generally diffuse outwards along the direction in which transport is easiest, which reduces the probability of multiple-scattering events and therefore narrows the broad component. Naturally, the models with optically thin spatial directions produce prominent narrow lines, because photons can escape without moving through the hot region. 

Ultimately, adding such angular geometric freedom results in i) a potential explanation for simultaneously seeing narrow components dominating the flux and broad many-scattering components ii) further uncertainties in estimating $\tau_{es}$ from the broad width. However, angular degrees of freedom in the positions of surrounding hot electrons do not naturally provide a means to produce two distinct scattering components, as seen in the data (i.e. Sect.~\ref{sec:two_comp_es}). Departures from spherical symmetry are likely intrinsic to LRNe \citep{Kaminski2018,Steinmetz2024}, but we generally throughout this manuscript have opted for a simpler spherical framework. This is observationally motivated in these epochs from the close coupling of nebular and NIR-stellar emission indicative of a singular process (see Sect.~\ref{sec:two_comp}), whereas two distinct emission-components with separate photon diffusion timescales likely require fine-tuning for comparable components with consistent, close coupling.

% ======= where you want the table =======
\begin{figure}[t]
    \includegraphics[width=\linewidth]{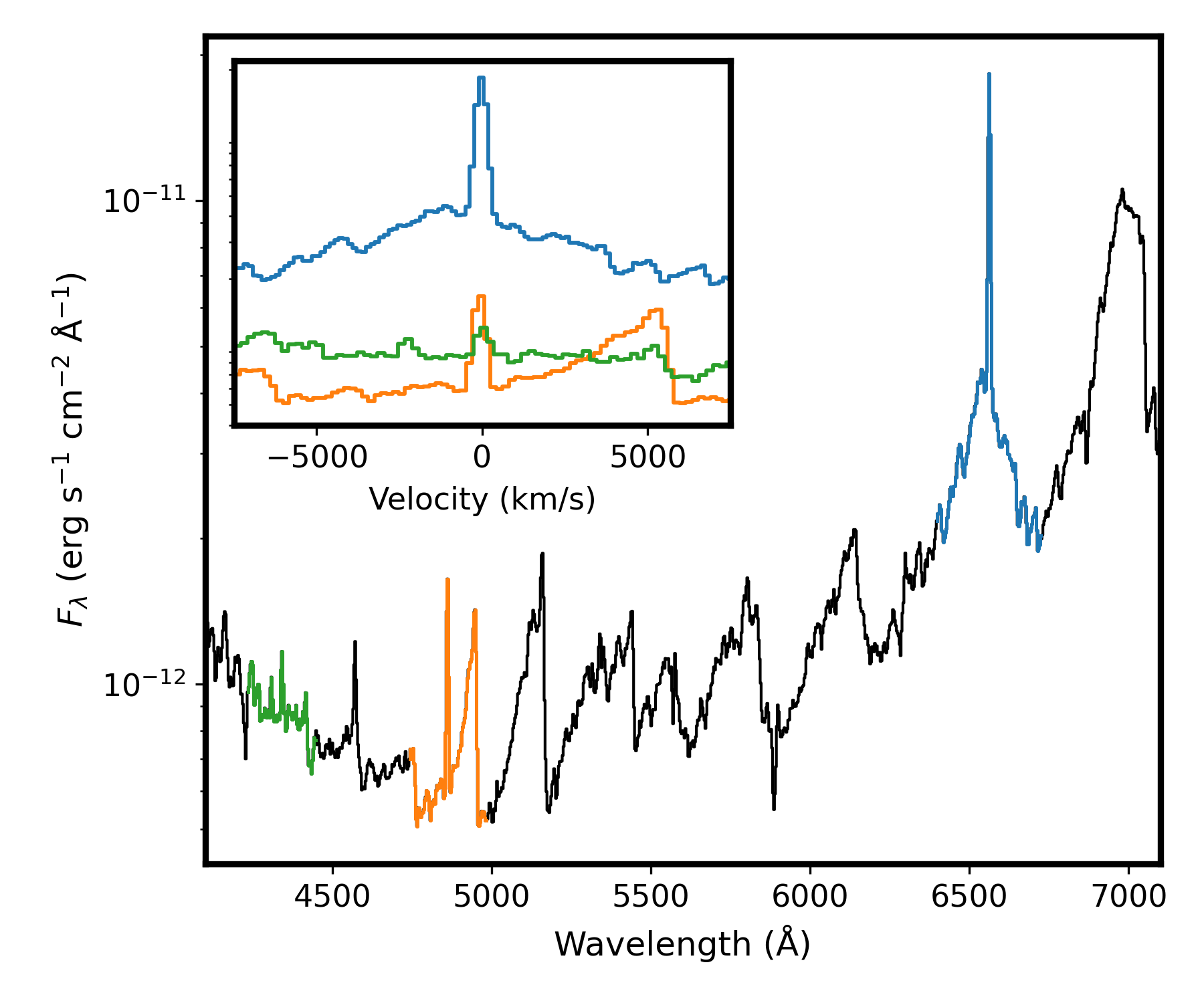}
    \caption{First post-plateau phase spectrum of V838 Mon (82 days after peak, 29/04/2002). Potentially a broad blueward-skewed exponential line is sitting beneath H$\alpha$, but this could (in contrast to the wings observed in Fig.~\ref{fig:Halpha_wind}) likely be explained the broad features induced by molecular bands. H$\beta$ and H$\gamma$ are suppressed below Case~B recombination ratios (H$\beta$/H$\alpha=0.051\pm0.001$, H$\gamma$/H$\beta=0.30\pm0.01$).}
    \label{fig:V838Mon}
\end{figure}

\begin{table*}
\caption{Summary of spectral diagnostics across LRNe. Columns indicate the presence of electron-scattering wings, the spectral range covered, \(E(B-V)\), whether Case~B recombination ratios are applicable.}
\setlength{\tabcolsep}{6pt}
\renewcommand{\arraystretch}{1.2}
\begin{tabularx}{\textwidth}{l X X X X X}
\toprule
\textbf{Object} &
\textbf{Electron-scattering} &
\textbf{Spectral range (nm)} & 
\(\mathbf{E(B-V)}\) &
\textbf{Case~B} &
\textbf{$L_{\mathrm{peak}}$ (\(\mathrm{erg\,s^{-1}}\))} \\
\midrule
V1309\,Sco & Observed & 310 - 5000 [1,2] & 0.65–0.84 [3]  & Yes & $6\cdot10^{38}$  \\
\midrule
V838 Mon & Tentative & 400-700 [4] & 0.87 [5] & No & $5\cdot10^{39}$ \\
\midrule
NGC4490-2011OT1 & Observed  & 370-950 [6] & 0-2 & No & $3\cdot10^{41}$ \\
\midrule
UGC12307-2013OT1 & Observed & 450-950 [6] & 0.22 (Galactic) & N/A & $3\cdot10^{41}$\\
\midrule
M101-2015OT1 & Tentative & 400-950 [7,8] & 0.008 (Galactic) & No & $3\cdot10^{40}$  \\
\midrule
AT2021blu & N/A & 370-950 [9]  & 0.02 (Galactic) & No &$10^{41}$ \\
\bottomrule
\end{tabularx}
\tablefoot{AT2021blu does not have spectroscopy during the post-plateau phase and V838 Mon is characterised by a very prominent continuum, so the existence of electron-scattering wings cannot be definitively determined. NGC4490-2011OT1 and UGC12307-2013=T1 have poorly constrained dust properties, so the line-ratio constraints on Case B recombination are therefore uncertain. 1: \cite{Mason2010}, 2: \cite{Rudy2025}, 3: \cite{Mason2022}, 4: \cite{Munari2002b}, 5: \cite{Munari2005}, 6: \cite{Pastorello2019}, 7: \cite{Blagorodnova2017}, 8: \cite{Goranskij2016}, 9: \cite{Pastorello2023} }
\label{tab:lrn_diagnostics}
\end{table*}

\begin{figure*}
    \includegraphics[width=0.33\linewidth]{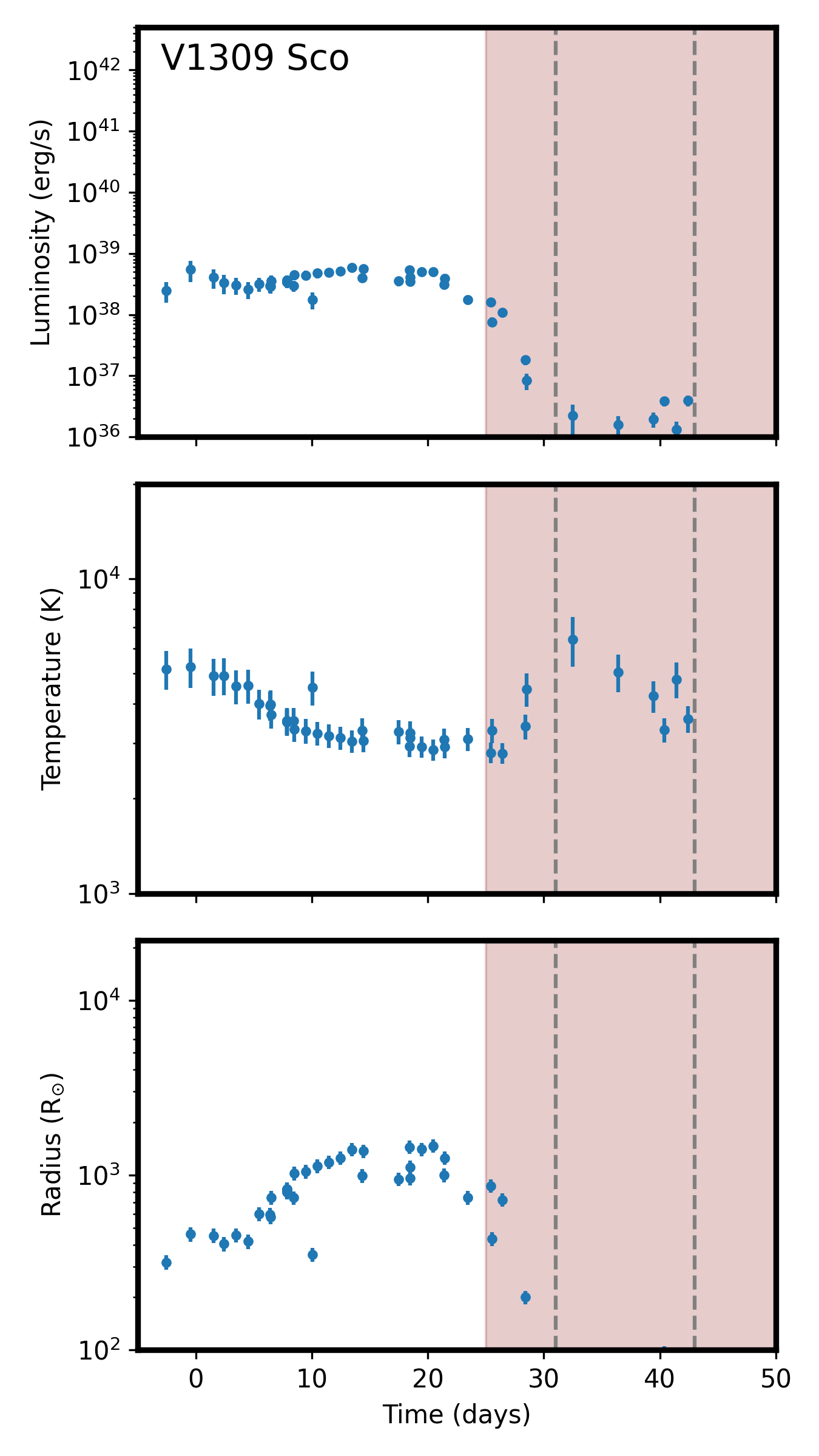}
    \includegraphics[width=0.33\linewidth]{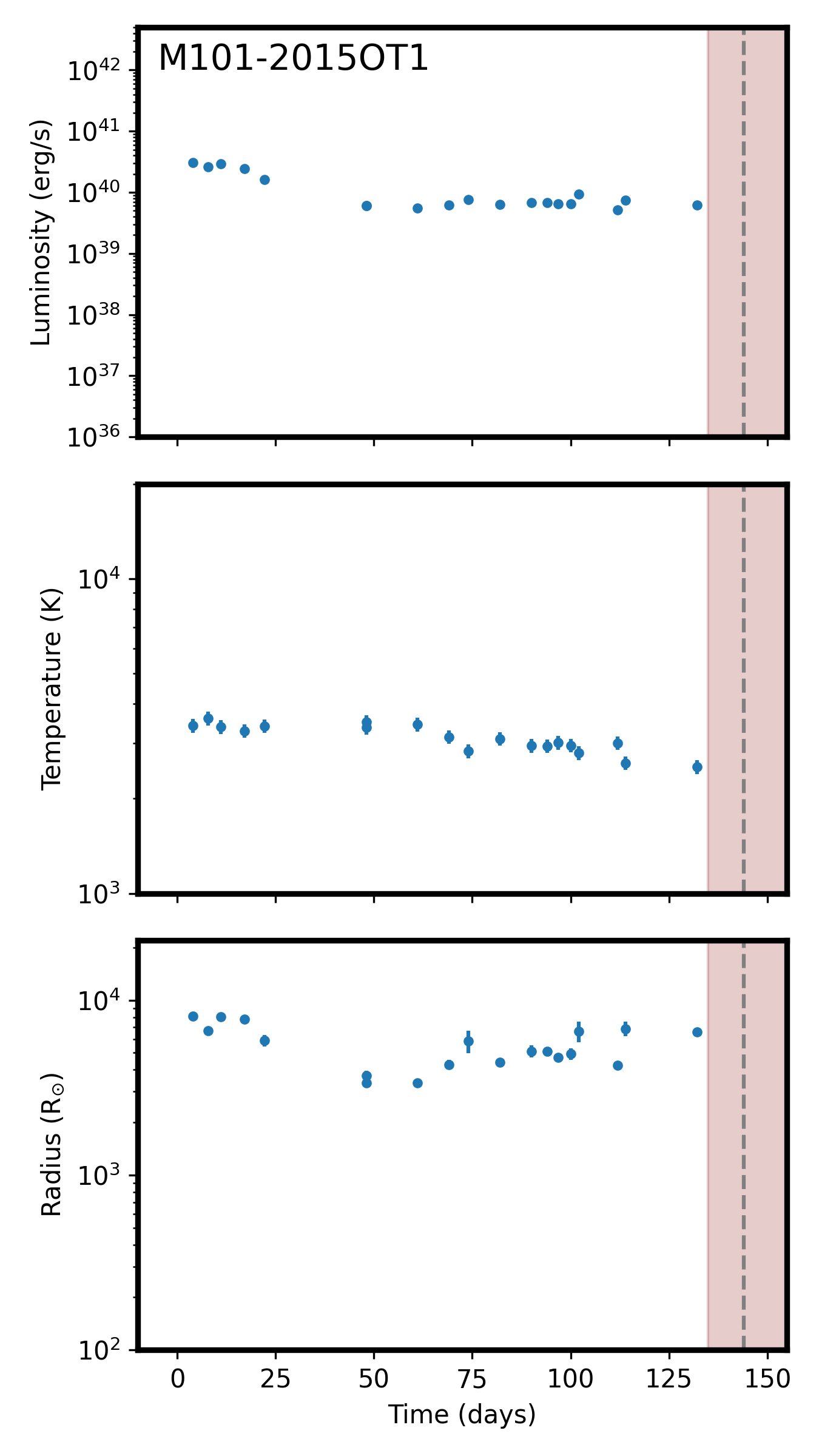}
    \includegraphics[width=0.33\linewidth]{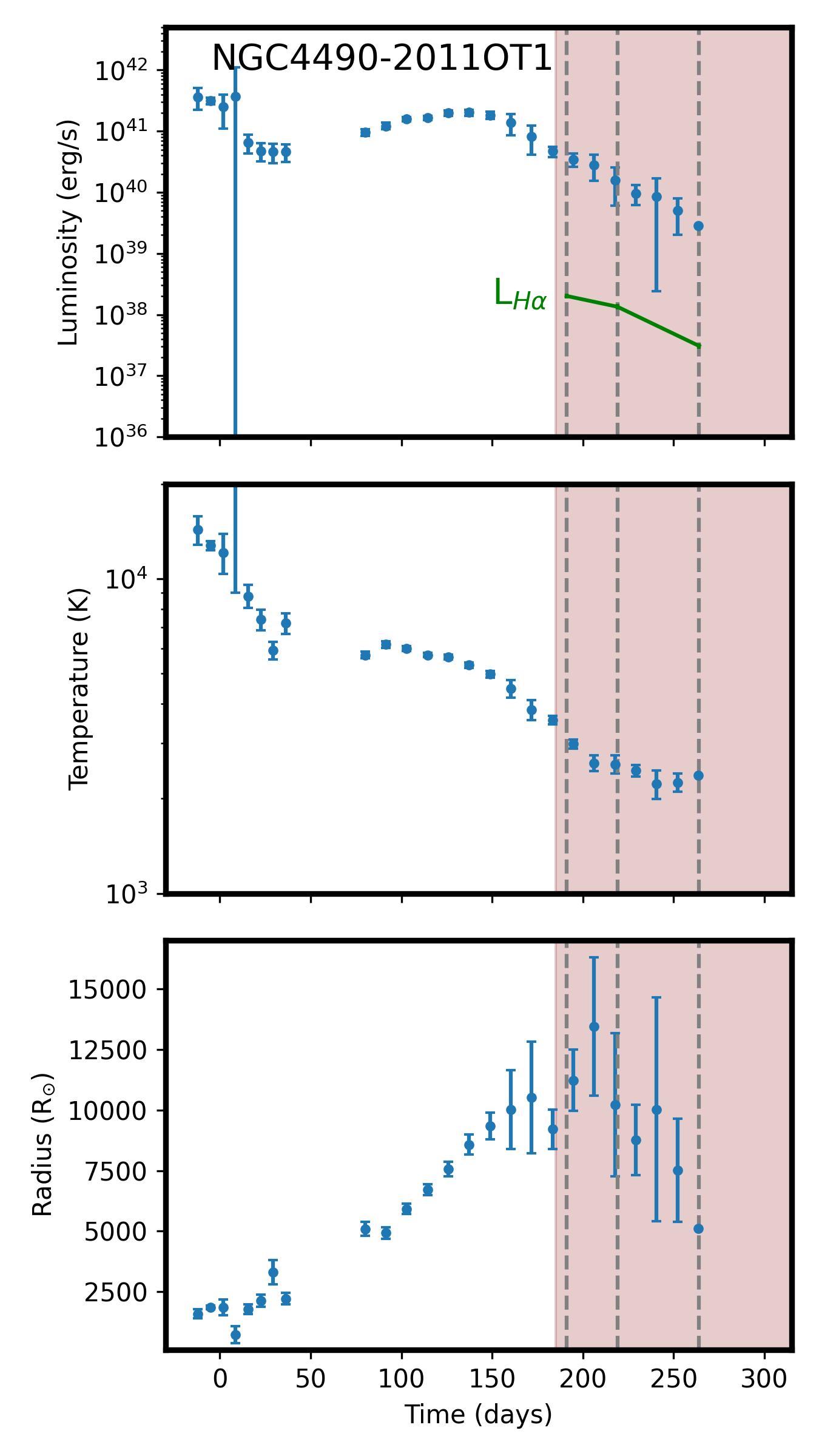}
    \includegraphics[width=0.33\linewidth]{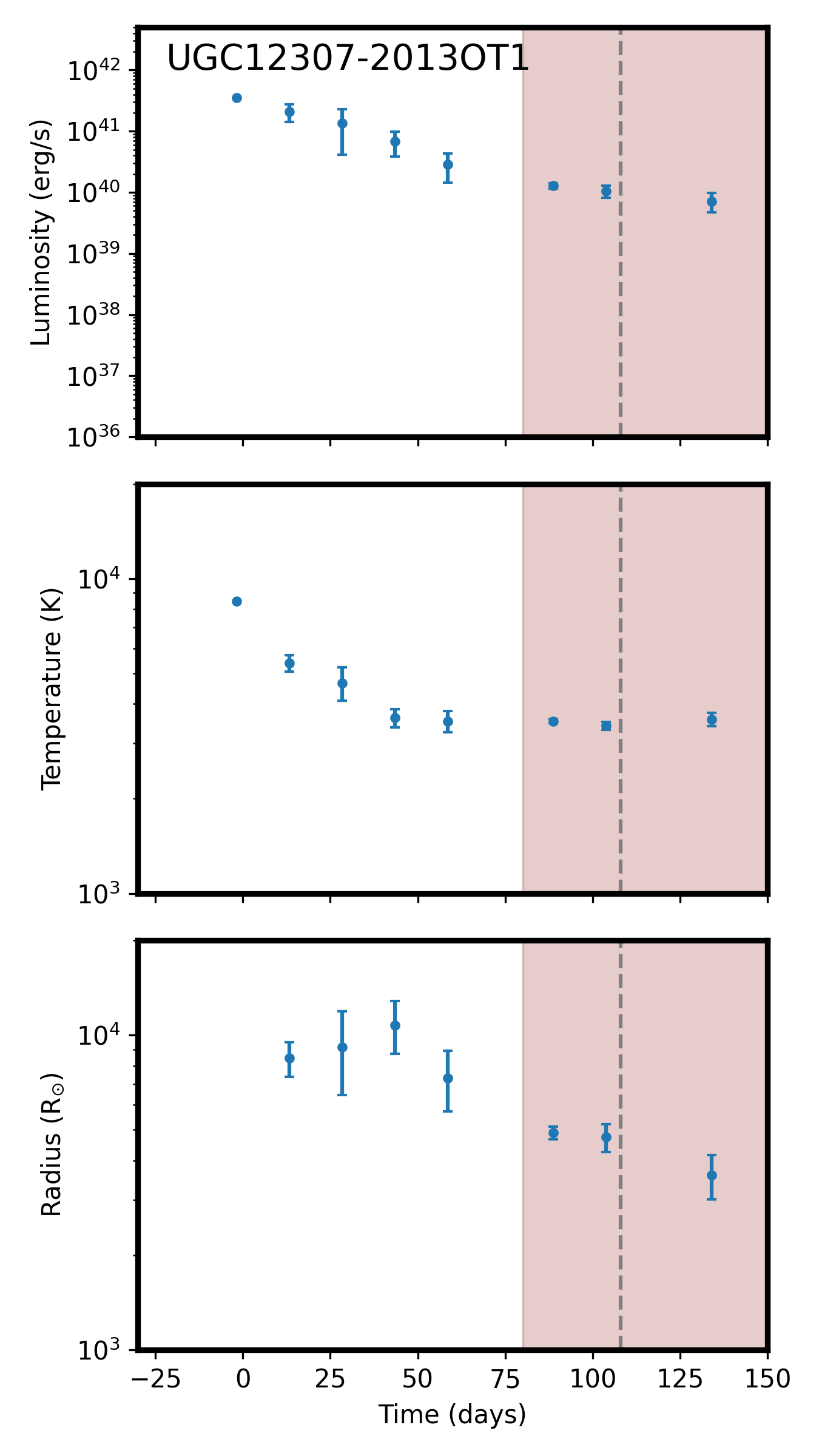}
    \includegraphics[width=0.33\linewidth]{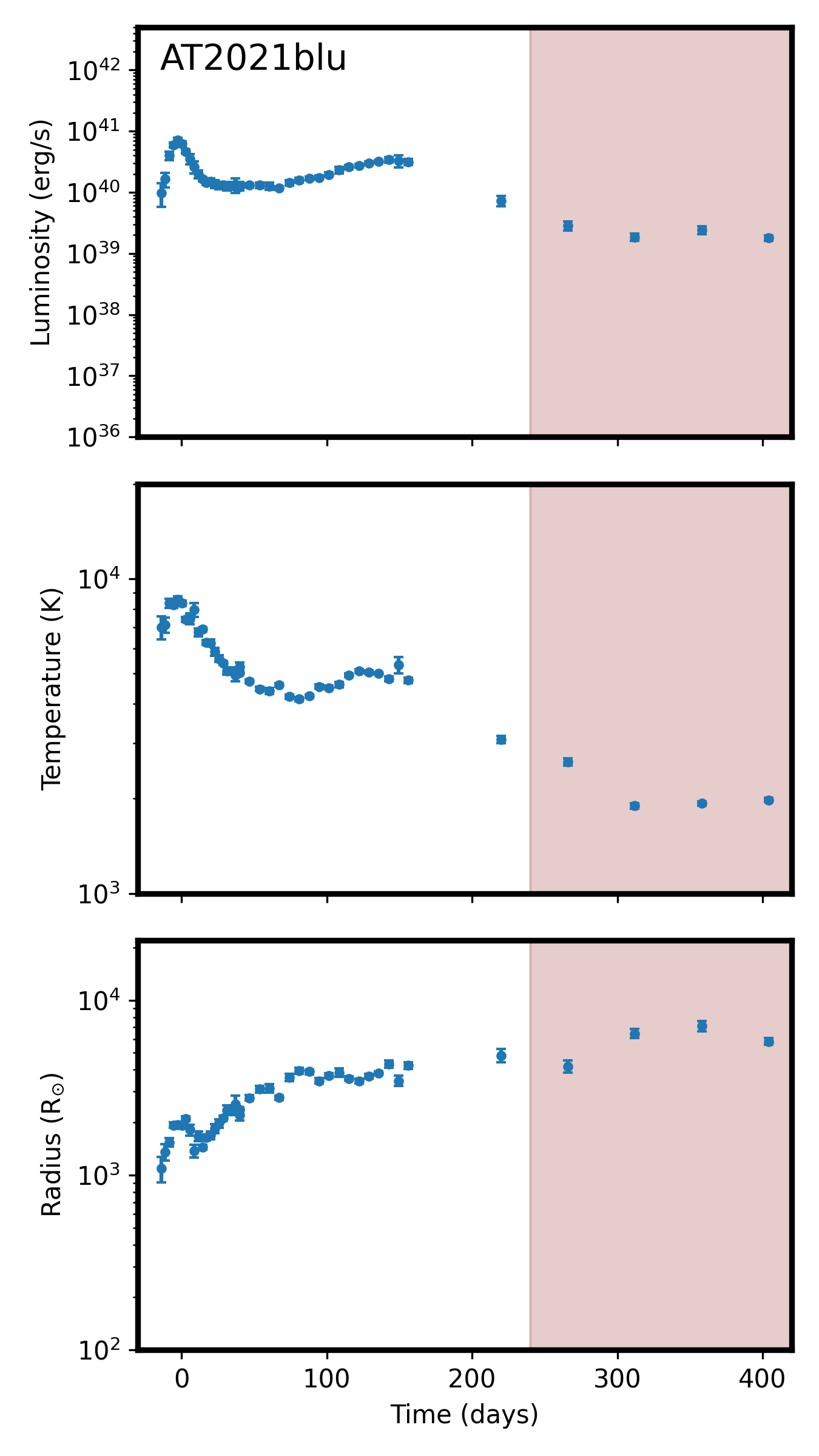}
    \includegraphics[width=0.33\linewidth]{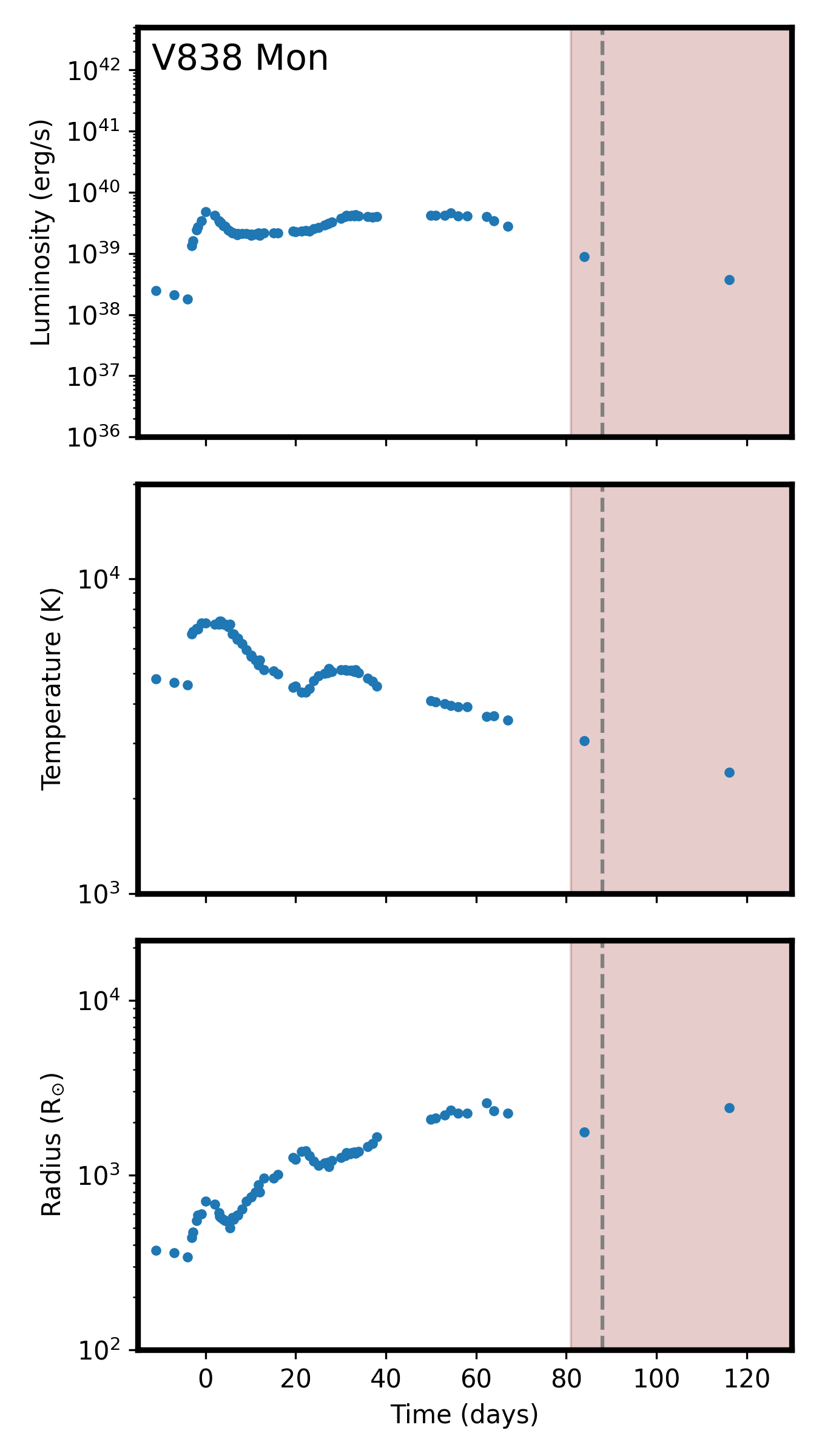}

    \caption{Example i) bolometric light curve, ii) best-fit blackbody temperature and iii) photospheric radius for 
    V1309\,Sco \citep{Tylenda2011}, M101-2015OT1 \citep{Blagorodnova2017}, NGC4490-2011OT1 and UGC12307-2013OT1 \citep{Pastorello2019}, AT2021blu \citep{Pastorello2023}, V838 Mon \citep{Tylenda2005}. Red region indicates the likely composite phase (as judged from the sudden deterioration of the blackbody fit when approaching $T\!\sim\!3000K$ limit), while the dashed grey lines indicate when post-plateau phase spectra where taken. These spectra contain broad H$\alpha$ lines (i.e. electron-scattering wings), although only tentatively for the earliest post-plateau phase NGC4490-2011OT1, M101-2015OT1 (due to low S/N, see Fig.~\ref{fig:Halpha_wind}) and for V838 Mon (due to a strong molecular band producing a coincident broad feature, \cite{Munari2002b}). For NGC4490-2011OT1 we also show the H$\alpha$ luminosity to emphasise the coherent fading of line and continuum.}
    \label{fig:lbol}
\end{figure*}

\section{Diversity of photospheric radii, blackbody temperatures, and spectral properties}\label{sec:BB_fit}
In Table~\ref{tab:lrn_diagnostics}, we provide a summary of the relevant spectral properties of these LRNe. 
In particular, we indicate which LRNe have (i) post-plateau phase spectroscopy showing electron-scattering wings, and (ii) constrained dust properties together with well-measured line ratios that show deviations from Case~B recombination due to the higher levels approaching their critical densities.
All the LRNe with early post-plateau spectroscopic data display broad features around H$\alpha$ (e.g. see Fig.~\ref{fig:Halpha_wind}), but as illustrated in Fig.~\ref{fig:V838Mon}, V838 Mon may be considered more ambiguous as large molecular features are present throughout the spectrum. 
Among published LRN spectra, V1309\,Sco is the only object we found with adequate S/N below 365\,nm. It is therefore the only object for which we can identify a Balmer break during the photospheric phase, and strong free-bound emission during the post-plateau phase.

In Fig.~\ref{fig:lbol}, we show the corresponding bolometric luminosity, temperature, and photospheric radii for these LRNe. These quantities are obtained with the photometric data being fit as a blackbody, $L_{\lambda}=4\pi R^2\pi B_{\lambda}(T)$. Here $B_{\lambda}(T)$ is the Planck function at temperature $T$, and the photospheric radius $R$ defines the emitting surface. When available we have corrected the photometry data for the Galactic contribution to reddening (see Table~\ref{tab:lrn_diagnostics}).  Reddening in some host-galaxies (particular NGC4490-2011OT1) may be poorly constrained, which can bias particularly the blackbody temperature. Most objects are well-sampled with 5-10 near-contemporaneous photometric bands, where blackbodies provide a good spectral approximation.
In general, the blackbody fitting framework is only meaningful for photospheric epochs, as post-plateau spectra are poorly described with a blackbody continuum. For V1309\,Sco, the constraints are provided solely by the optical bands. This means that, after the photospheric phase, a significant fraction of the flux is omitted, since the dominant emission shifts to the near- and mid-infrared \citep[e.g. ][]{Rudy2025}. In photospheric epochs, recombination lines provide a negligible contribution to the total luminosity, whereas in the post-plateau epochs these become a substantial fraction.

\begin{figure*}[t]
    \includegraphics[width=\linewidth]{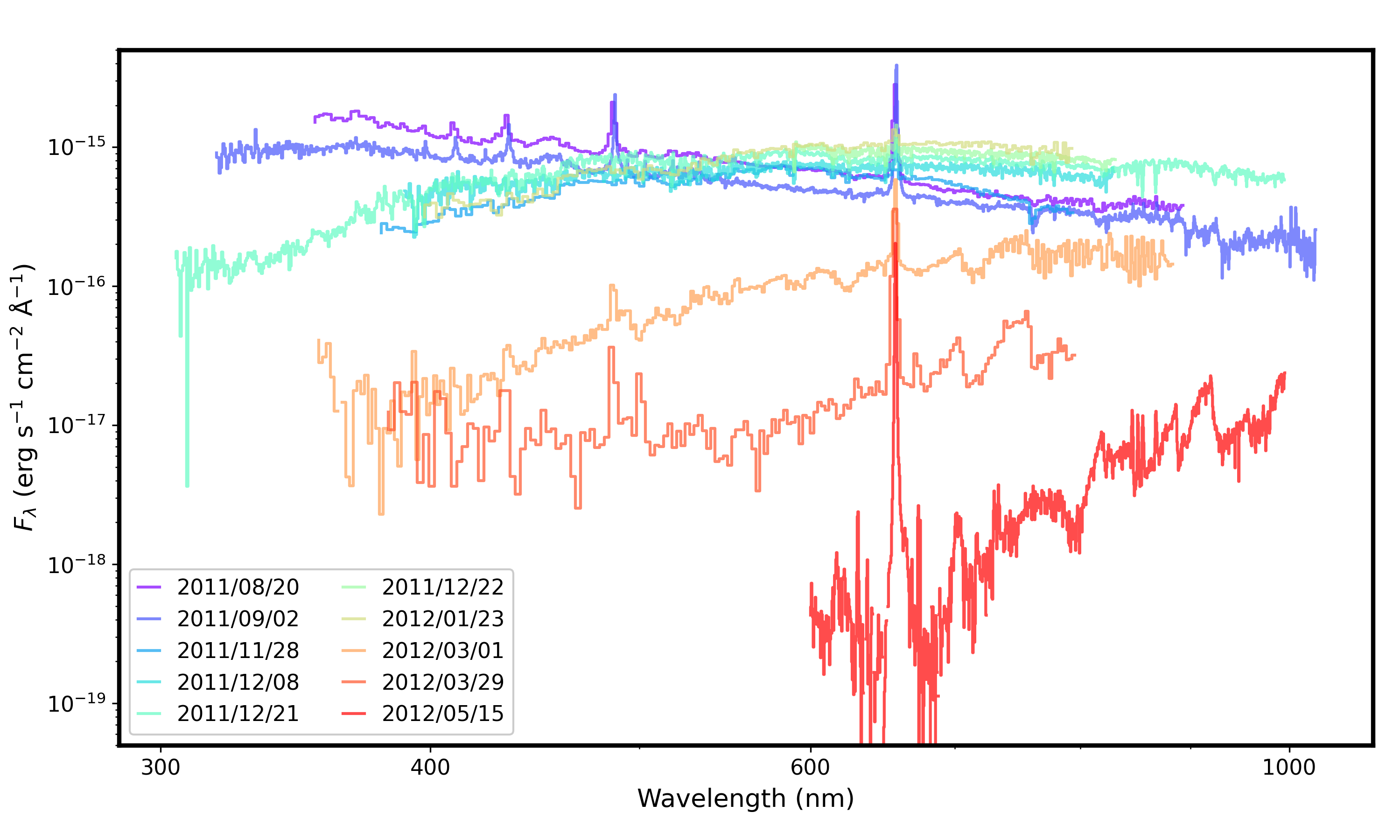}
    \caption{ Spectral series of NGC4490-2011OT1 \citep{Pastorello2019} including post-plateau epochs (red), where i) prominent hydrogen recombination lines (H$\alpha$ and Paschen series) and ii) broad electron-scattering wings are observed. }
    \label{fig:NGC4490}
\end{figure*}

\end{document}